\documentclass[%
 aip,
 amsmath,amssymb,
 reprint,%
]{revtex4-1}

\usepackage{graphicx}
\usepackage{dcolumn}
\usepackage{bm}

\usepackage[utf8]{inputenc}
\usepackage[english]{babel}
\usepackage[T1]{fontenc}
\usepackage{mathptmx}
\usepackage{float}
\usepackage{xcolor}
\usepackage{multirow}

\definecolor{amaranth}{rgb}{0.9, 0.17, 0.31}

\definecolor{blue}{rgb}{0.0, 0.3, 1}

\definecolor{green}{rgb}{0.0, 0.5, 0}

\begin{document}

\preprint{AIP/123-QED}

\title{Dynamics in cortical activity revealed by resting-state MEG rhythms}

\author{J. Mendoza-Ruiz}
 \affiliation{Department of Statistics, Universidad Nacional, Cr 45 $\#$ 26-85, Bogot\'a, Colombia.}

\author{C. E. Alonso-Malaver}
 \affiliation{Department of Statistics, Universidad Nacional, Cr 45 $\#$ 26-85, Bogot\'a, Colombia.}
 
\author{M. Valderrama}
\affiliation{Biomedical Engineering Department, Universidad de los Andes, Cr 1 $\#$ 18A-12, Bogot\'a, Colombia.}

\author{O. A. Rosso}
\affiliation{Instituto de F\'{\i}sica, Universidade Federal de Alagoas (UFAL), BR 104 Norte km 97, 57072-970, Macei\'o, Alagoas, Brasil.}

\author{J.H. Martinez}
\email{johemart@gmail.com}
\affiliation{Biomedical Engineering Department, Universidad de los Andes, Cr 1 $\#$ 18A-12, Bogot\'a, Colombia.}

\affiliation{Grupo Interdisciplinar de Sistemas Complejos (GISC), Madrid, Spain.}

\date{\today}

\begin{abstract}
The brain is a biophysical system subject to information flows that may be thought of as a many-body architecture with a spatio-temporal dynamics described by its neuronal structures. 
The oscillatory nature of brain activity allows these structures (nodes) to be described as a set of coupled oscillators forming a network where the node dynamics, and that of the network topology can be studied.
Quantifying its dynamics at various scales is an issue that claims to be explored for several brain activities, e.g., activity at rest. 
The resting-state (RS) associates the underlying brain dynamics of healthy subjects that are not actively compromised with sensory or cognitive processes. Studying its dynamics is highly non-trivial but opens the door to understand the general principles of brain functioning, as well as to contrast a passive null condition versus the dynamics of pathologies or non-resting activities.
Here we hypothesize about how could be the spatio-temporal dynamics of cortical fluctuations for healthy subjects at RS. 
To do that, we retrieve the alphabet that reconstructs the dynamics (entropy/complexity) of magnetoencephalograpy (MEG) signals.
We assemble the cortical connectivity to elicit the dynamics in the network topology. 
We depict an order relation between entropy and complexity for frequency bands that is ubiquitous for different temporal scales.
We unveiled that the posterior cortex conglomerates nodes with both stronger dynamics and high clustering for $\alpha$ band.
The existence of an order relation between dynamic properties suggests an emergent phenomenon characteristic of each band.
Interestingly, we find the posterior cortex as a domain of dual character that plays a cardinal role in both the dynamics and structure regarding the activity at rest.
To the best of our knowledge, this is the first study with MEG involving information theory and network science to better understand the dynamics and structure of brain activity at rest for different bands and scales.
\end{abstract}

\maketitle

\begin{quotation}
Studying the RS dynamics is highly non-trivial but opens the door to understand the general principles of brain functioning. A relevant question is how much information the cortical fluctuations convey among neural structures.
Entropy and complexity are candidates to evaluate the information content of MEG signals of healthy subjects.
We propose to capture the dynamics of cortical structures and that of the functional network at different bands for several time-scales, via ordinal patterns and clustering coefficient.
We evidence an order relation between entropy and complexity regarding brain rhythms.
We unveil the posterior cortex as the one that conglomerates structures with high levels of both dynamics and clustering.
Our results confirm the emergence of certain information processing typical of each band with topographical localization at the occipital lobe.
\end{quotation}

\section{\label{sec:intro} Introduction}
The human brain is probably one of the most complex systems we face, being this main reason for its study and fascination about its dynamics and how different activity organization is carried out. At present days, we could say that the structure of the human brain (anatomy) is well known, however, as far as, its dynamics and how it creates thoughts, processes emotions, and perceptions, it is not fully understood. The development of new image acquisition and processing technologies have made it possible to look inside a living brain and see it at work. The brain works by generating small electrical imbalances in neural membranes. Functional images reveal which areas are most active. This can be done by directly measuring electrical activity (EEG), by capturing magnetic fields created by electrical activity (MEG), or by measuring metabolic side effects such as alterations in glucose uptake (PET) and blood flow (fMRI).\\
The brain is composed of the cerebrum, cerebellum, and brainstem ~\cite{Carter2019}. The cerebrum is the dominant part of the brain and, it is the large pinky-gray wrinkled structure that forms more than three-quarters of the brain's total volume. The cerebrum is divided into left and right hemispheres, which are linked by a ``bridge'' of nerve fibers, the corpus callosum. The cerebellum is located under the cerebrum, its main function is to coordinate muscle movement,  maintain posture, and balance. 
The brain's physical structure broadly reflects its mental organization. In general, higher mental processes occur in the upper regions, while the brain's lower regions take care of basic life support ~\cite{Carter2019}.\\
The uppermost brain region, the cerebral cortex, is mostly involved in conscious sensations, abstract thought processes, reasoning, planning, working memory, and similar higher mental processes. The limbic areas on the brain's innermost sides, around the brainstem, deal largely with more emotional and instinctive behaviors and reactions, as well as long-term memory. The thalamus is a preprocessing and relay center, primarily for sensory information coming from lower in the brainstem, bound for the cerebral hemispheres above. Moving down the brainstem into the medulla are the so-called ``vegetative'' centers of the brain, which sustain life even if the person has lost consciousness. The brain's vertical zonation moves from high-level mental activity in the cerebral cortex gradually through to more basic or ``primitive'' lower functions, especially the autonomic centers of the medulla in the lower brainstem that deal with vital body functions, such as breathing and heartbeat ~\cite{Carter2019}.\\
Structurally, the left and right cerebral hemispheres look broadly similar. Functionally, however, speech and language, stepwise reasoning and analysis, and certain communicating actions are based mainly on the left side in most people.\\
Since nerve fibers cross from left to right at the base of the brain, this dominant left side receives sensory information from and sends messages to, muscles on the right side of the body.
Meanwhile, the right hemisphere is more concerned with sensory inputs, auditory and visual awareness, creative abilities, and spatial-temporal awareness ~\cite{Carter2019}.\\  
The cerebral cortex is the outer layer of the brain's most dominant part, the cerebrum. It is the bulging wrinkled surface we see when looking at the brain from any angle. It is commonly known as gray matter from its color, which contrasts with the white matter in the layer below. Bulges and grooves help divide the cortex into four paired lobes: frontal, temporal, parietal, and occipital. The main and deepest groove is the longitudinal fissure that separates the cerebral hemispheres. The highly convoluted sheet of gray matter that constitutes the cerebral cortex varies in thickness from about 2 $mm$ to 5 $mm$. Estimates of its cell numbers vary from $10 \time 10^9$ to more than $50 \times 10^9$ neurons and about 5 to 10 times this number of glial (supporting) and other cells ~\cite{Carter2019}.\\
The human cortex contains a distinct pattern of neuron types. 
Cortical neurons receive and send signals to other brain areas, including other parts of the cortex. 
This to and from of messages keeps all parts of the brain aware of what is going on elsewhere. Neurons in the cortex are ``head down'' -- their receiving parts (dendrites) point up to the surface, while threads that carry messages to other cells (axons) are oriented down. Some axons extend below the cortex and form part of the ``white matter'' connective tissue that carries information to distant brain areas. Other axons travel through the lower layers of the cortex to connect with other cortical cells ~\cite{Carter2019}.\\
There are over a thousand types of brain cells, which fall into two broad groups: neurons and glial cells. Neurons send electrical signals, or ``fire'', in response to stimuli. There are about $86 \times 10^9$ neurons in an average human brain and ten times as many glial cells. Neurons can be categorized structurally according to the location of the cell body in relation to the axon and dendrites, and also the number of dendrites and axon branches. In the cortex, one neuron may receive signals from many thousands of other neurons via its multitudinous branching dendrites. Signals are conducted to the soma, around this, and then away along the axon--always by the cell membrane.
Glial cells give physical support to neurons, but they are also thought to influence neurons' electrical activity. They provide physical support for the thin dendrites and axons that wind their way around the neural network and supply nutrition for neurons in the form of sugars and raw materials for growth and repair ~\cite{Carter2019}.
Synapses are communication sites where neurons pass nerve impulses among themselves. Many neurons do not actually touch one another, but pass their signals via chemicals (neurotransmitters) across an incredibly thin gap, called the synaptic cleft. Synapses are divided into types according to the sites where the neurons almost touch, e.g., the soma, the dendrites, the axons, and tiny narrow projections called dendritic spines ~\cite{Carter2019}.\\
A nerve impulse or signal can be thought of as a tiny, brief ``spike'' of electricity traveling through a neuron. At a more fundamental level, it consists of chemical particles moving across the cell's outer membrane, from one side to the other. 
Nerve signals are composed of a series of discrete impulses, also known as action potentials. A single impulse is caused by a traveling ``wave'' of chemical particles called ions, which have electrical charges and are mainly the minerals sodium, potassium, and chloride. In the brain, and throughout the body, most impulses in most neurons are of the same strength, about 100 $mV$. They are also of the same duration, around one millisecond, but travel at varying speeds. The information they convey depends on how frequently they pass in terms of impulses per second, where they came from, and where they are heading ~\cite{Carter2019}.\\
In the normal brain, an action potential travels down the axon to the nerve terminal where a neurotransmitter is released. At the postsynaptic membrane, the neurotransmitter produces a change in the membrane conductance and transmembrane potential. If the signal has an excitatory effect on the neuron it leads to a local reduction of the transmembrane potential (depolarization) and it is called an excitatory postsynaptic potential (EPSP), typically located in the dendrites. If the signal has an inhibitory effect on the neuron it leads to local hyperpolarization, also called an inhibitory postsynaptic potential (IPSP), typically located on the cell body of the neuron. The combination of EPSPs and IPSPs induces currents that flow within and around the neuron with a potential field sufficient to be recorded on the scalp. These ionic currents are in the range of nano ampere ($10^{-9}A$). When it circulates, generate magnetic induction that can be measured on the scale of femtoteslas ($10^{-15}T$) ~\cite{Carter2019}.\\
The electroencephalography (EEG) ~\cite{Jackson2014,Cohen2017} and the magnetoencephalography (MEG) ~\cite{Hamalainen1993,Baillet2017} are the most extended non-invasive physiologic techniques that allow one to measure electrical and magnetic activity generated by the brain on the scalp surface. 
EEG and MEG are different but complementary techniques recorded at the scalp for observing brain electromagnetic activity (see the review by Sylain Baillet ~\cite{Baillet2017}). \\
On the one hand, EEG measures the electrical activity in order of microvolts ($\mu V$), due to the difference of extracellular potential from cortical groups of pyramidal neurons. On the other hand, MEG measures the magnetic activity, in the scale of femtoteslas ($fT$) of large axons in the same group of neurons. These signals recorded on the scalp have a poor relationship with the spiking activity of individual neurons. They are spatial-temporally smoothed versions of cortical neuronal activity under an area of approximately $10~\text{cm}^2$ and, can accurately detect brain activity at the time resolution of a single millisecond (ms). Therefore, mental states would emerge from the dynamical interaction between multiple physical and functional levels, giving origin to the oscillatory rhythmic brain activity. They are of functional importance to understand how information is processed in the brain, highlighting oscillations bands like: 
delta $\delta \in [0.5-4) \ Hz$, 
$\theta \in (4-7)\ Hz$, $\alpha \in (8-13)\ Hz$, $\beta \in (14-29)\ Hz$ and, $\gamma \geq 30\ Hz$ ~\cite{Herrmann2016}.\\
If we have an observational set of measures (MEG time series, ${\mathcal  X}(t)$) from a dynamical system whose evolution can be tracked through time, a natural question arises: How much information these observations encoding about the dynamics of the underlying system (i.e., the brain dynamics). We adopted an Information Theory point of view ~\cite{Gray2011}. The information content of the system is typically evaluated via a probability distribution function (PDF) $P$ describing the apportionment of the observable quantity (the time series ${\mathcal  X}(t)$). Quantifying the information content of a given observable is therefore largely tantamount to characterizing its PDF. This is often done with a wide family of measures called Information Theory Quantifiers (ITQ). The ITQ can characterize relevant properties of the PDF associated with the time series, and represent metrics of the space of PDF's for the data set, allowing to compare different sets and classify them according to the properties of the underlying processes (deterministic vs stochastic) ~\cite{PRL-Rosso2007}.  \\
The temporal brain dynamics is our focus of interest and, the measured data are the time series  ${\mathcal  X}(t)$ of MEG. Metrics that take the temporal order of observations explicitly into account are of our interest. That is, our approach is fundamentally a ``causal" (the data sequence determines the PDF) rather than ``statistical" one (the correlation between successive values are destroyed or not consider in the construction of the PDF). We follow the Bandt and Pompe methodology ~\cite{Bandt-Pompe2002} which is a simple and robust symbolic procedure that takes into account the time causality of ${\mathcal  X}(t)$ (causal coarse-grained methodology) by comparing neighboring values in the time series (permutation BP-PDF).\\
The ITQ selected are the {\it Shannon Entropy}, ${S}[P]$ ~\cite{Shannon1948,Weaver1949} and, the effective {\it Statistical Complexity Measure}, ${C}[P]$ ~\cite{Lamberti2004}, which will be evaluated using permutation BP-PDF. Entropy is a basic quantity with multiple field-specific interpretations; for instance, it has been associated with the disorder, state-space volume, and lack of information ~\cite{Shannon1948,Weaver1949}. When dealing with information content, the Shannon entropy is often considered as the foundational and most natural one ~\cite{Brissaud2005}. In contrast to information content, there is not a universal definition of complexity. Between two special instances of perfect order and high entropy, a wide range of possible degrees of physical structure exists that should be reflected in the features of the underlying probability distribution $P$. One would like to assume that the degree of correlational structures would be adequately captured by some functional ${C}[P]$ in the same way that Shannon's entropy  ${S}[P]$ ~\cite{Shannon1948} ``captures" randomness. The ordinal structures present in the process are not quantified by randomness measures, and consequently, measures of structural complexity are necessary for a better understanding (characterization) of the system dynamics represented by their time series ~\cite{Crutchfield1998}. The opposite extremes of perfect order and maximal randomness are very simple to describe because they do not have any structure. The complexity should be zero in these cases. At a given distance from these extremes, a wide range of possible ordinal structures exists.\\
Complexity can be characterized by a certain degree of organization, structure, memory, regularity, symmetry, and patterns ~\cite{Feldman2008}. The complexity measure does much more than satisfy the boundary conditions of vanishing in the high- and low-entropy limits. In particular, the maximum complexity occurs in the region between the system's perfectly ordered state and the perfectly disordered one. Complexity allows us to detect essential details of the dynamics, and more importantly to characterize the correlational structures of the orderings present in the time series. A suitable measure of complexity can be defined as the product of a measure of information and a measure of disequilibrium (i.e., some kind of distance from the equilibrium PDF $P_e$ to the accessible actual states of the system $P$) ~\cite{LMC1995}. In particular, Rosso and coworkers ~\cite{PRL-Rosso2007} introduced an effective Statistical Complexity Measure (SCM) ${C}[P]$, that can detect essential details of the underlying dynamical processes.\\
Dynamics is of high relevance, but more recently, the structure and function of the brain have begun to be investigated using ``Network Science'' ~\cite{Costa2007}, offering a large number of quantitative tools and properties, thus greatly enriching the set of objective descriptors of brain structure and function available to neuroscientists ~\cite{Bullmore2009,Rubinov2010}. The link between ``network science'' and ``neuroscience'' has shed light on how the entangled anatomy of the brain is, and how cortical activations may synchronize to generate the so-called functional brain networks. Within this context, complexity appears to be the bridge between the topological and dynamical properties of biological systems and more specifically, the interplay between the organization and dynamics of functional networks. Particularly, how cortical activations can be understood as an output of a network of dynamical systems that are intimately related to the processes occurring in the brain.\\
Early neuroimaging and electrophysiological studies were usually aimed at identifying task-specific areas of activation, as well as local patterns varying over time. Nowadays, there exists a common knowledge that task-related brain activity is temporally multiscale and spatially extended, in the same way as networks of coordinated brain areas are continuously formed and destroyed. The studies of functional brain activity have focused on identifying the specific nodes forming these networks and on characterizing the metrics of connectivity between them. The underlying hidden hypothesis being that each node, which constitutes a coarse-grained representation of a given Region of Interest (ROI), makes a unique contribution to the whole. In this way, functional neuroimaging initially integrated the two basic ingredients of early neuropsychology: the localization of cognitive functions in specialized brain modules and the role of connection fibers in the integration of them. \\
From the structural point of view, the human brain can be understood as a network of cells forming a massively parallel system, organized to carry out three major functions: \textit{a)} computation, \textit{b)} information storage and transport, and \textit{c)} communication among computational structures. The brain archives impressively high levels of computational resources by adopting efficient architectures, involving the timing of signals and the representation of information with energy-efficient codes, distributing signals appropriately in space and time. Cortical activations allow being studied in a two-face direction described above: by looking at the dynamics of cortical signals, or by taking into account the dynamics occurring in the brain connectivity. Brain dynamics characterization has been used to explore the reorganization of networks in mild cognitive impairment  ~\cite{Buldu2011}; the modularity of connectivity patterns in epilepsy brain networks and normal subjects ~\cite{Chavez2010}; 
the effects of memory in brain networks in old and young individuals, the interchange of information between brain hemispheres in resting-state  ~\cite{Johann2018SicRep1}; 
the characterization of visuomotor/imaginary movement in EEG ~\cite{Rosso2019Frontiers}; to name a few. From the perspective of ITQ, they were used to characterize and classify EEG records from control and epileptic patients ~\cite{Rosso2006JNM,Rosso2005PhysA,Rosso2009JNM1,Rosso2009JNM2,Rosso2017Entropy};
it has also been applied to differentiate processing information zones for subjects with Alzheimer at several frequency bands ~\cite{Echegoyen2020}; 
to discriminate imagined and non-imagined tasks in motor cortex area and its relation with rhythmic oscillations ~\cite{Rosso2018PhysA,Rosso2018Chaos,Rosso2018Entropy}; to evidence the irreversibility aspect of EEG at resting-state ~\cite{Zanin2020}  and epilepsy ~\cite{Johann2018chaos}; 
to unveil a relationship between the dynamics of electrophysiological signals and the brain network structure ~\cite{Johann2018SciRep2}; 
and even to propose a new ordinal-structure methodology to better account for the information transit between brain signals  ~\cite{Echegoyen2019}.\\
While the applicability of these methodologies spans over a wide range of neural phenomena, these applications are also of importance when concerning the activation of healthy subjects that are not actively compromised with the sensory or cognitive processes. This fact allows to contrast results from pathologies or non-resting activities versus, a passive null condition also called \textit{ Resting-State} (RS). The RS becomes useful to observe the underlying brain dynamics under normal M/EEG, specifically for $\alpha$ rhythm, which accounts for the range of frequencies with the highest energy in the spectrum ~\cite{Stam2005,Barry2007,Khanna2015,Masoller2018}. 
The dynamics and structure of RS have been explored all over the last decade. However, the relation between dynamics and topology of the brain at rest is still poorly understood.
It is in this scenario where this work lies. We hypothesized about how could be the information content of the MEG time series for a healthy group of subjects at RS. Specifically, how the spatio-temporal complexity of cortical fluctuations behaves.
To do that, we captured the dynamical properties of oscillations at the level of ROIs. We measure the dynamical properties of both the entropy and complexity through symbolic representation and computed structural features extracted from the connectivity pattern. We evidenced entropy-complexity correlations at different temporal scales. 
Our results depict an order relation among dynamical parameters for several frequency bands.
We unveiled the occipital lobe as the one with higher levels of complexity in the $\alpha$ band. 
We detected cortical regions of high levels of network clustering around the same occipital lobe. 
This fact highlights the role of occipital lobe where cortical regions influence both the dynamics (entropy/complexity) and the structure (clustering).\\
Our manuscript is organized as follows: 
Section II describes the MEG data set from the \textit{Human Connectome Project}.
Section III is divided into three parts. The first accounts for the extraction of the information content from cortical ${\mathcal  X}(t)$, the second is about the computation of the permutation entropy ($H$) and statistical complexity ($C$) for ${\mathcal  X}(t)$, the last one for describing the statistical coherence to gather functional networks, and network properties, i.e., clustering ($c_w$). 
Section IV introduces the results of the multiscale analysis in several frequency bands, with special attention on the $\alpha$ rhythm.
In Section V we related the results with structure and dynamics. 
Section VI is for conclusions,
\section{\label{sec:dataset} Dataset}
Data set consists of MEG time series ${\mathcal  X}(t)$. 
In this preliminary study, 190 magnetometers were located in the scalps of 40 healthy adult subjects (21 male, 19 female).
The range of age was chosen to represent adults beyond the age of major neurodevelopmental changes, and before the onset of neurodegenerative conditions.
Cortical signals were collected using a MAGNES 3600 system (4D Neuroimaging San Diego) housed in a magnetically shielded room, with subjects laying down with open eyes, instructed to relax, and fixation maintained on a projected red crosshair on a dark background. 
Magnetic field measures were taken in empty rooms to avoid excessive environmental noise and to estimate noise base levels. 
Data preprocessing was conducted to evaluate signals accuracy, identify low-quality sensors, avoid sharp increases in noise levels, and to confirm rough linearity in the range of sampling frequency. 
Correlations among neighbor sensors, variance ratio, and z-score values were conducted to detect flawed sensors. 
The procedure allowed us to identify some low-quality channels, which were then used in an Iterative Independent Component Analysis to identify other low-quality signals. The cardiac, ocular, and muscular activity were filtered from brain signals to remove artifacts. 
20 consecutive trials of 1018 samples each, were taken shaping the time series for all individuals. 
Each ${\mathcal  X}(t)$ is of $M= 20360$ length with sample frequency of 508.6 $Hz$ ~\cite{hcpmanual2014}. The experiment is reported in the Human Connectome Project. See detailed explanations in Van Essen's,  and Larson's work ~\cite{van2012,larson2013}. 
We tested weakly stationarity of signals with the Augmented Dicky-Fuller routine ~\cite{fuller2009} before using the band-pass filter ~\cite{cohen2014} to gather the bands ($\theta$, $\alpha$, $\beta$, and 
$\gamma$). All computations were performed with the statistical software $R$. 3.6.1.

\section{\label{sec:methods} Methods}
\subsubsection{\label{BP-PDF} Bandt-Pompe methodology for PDF}
\begin{figure}
\includegraphics[width=0.48\textwidth]{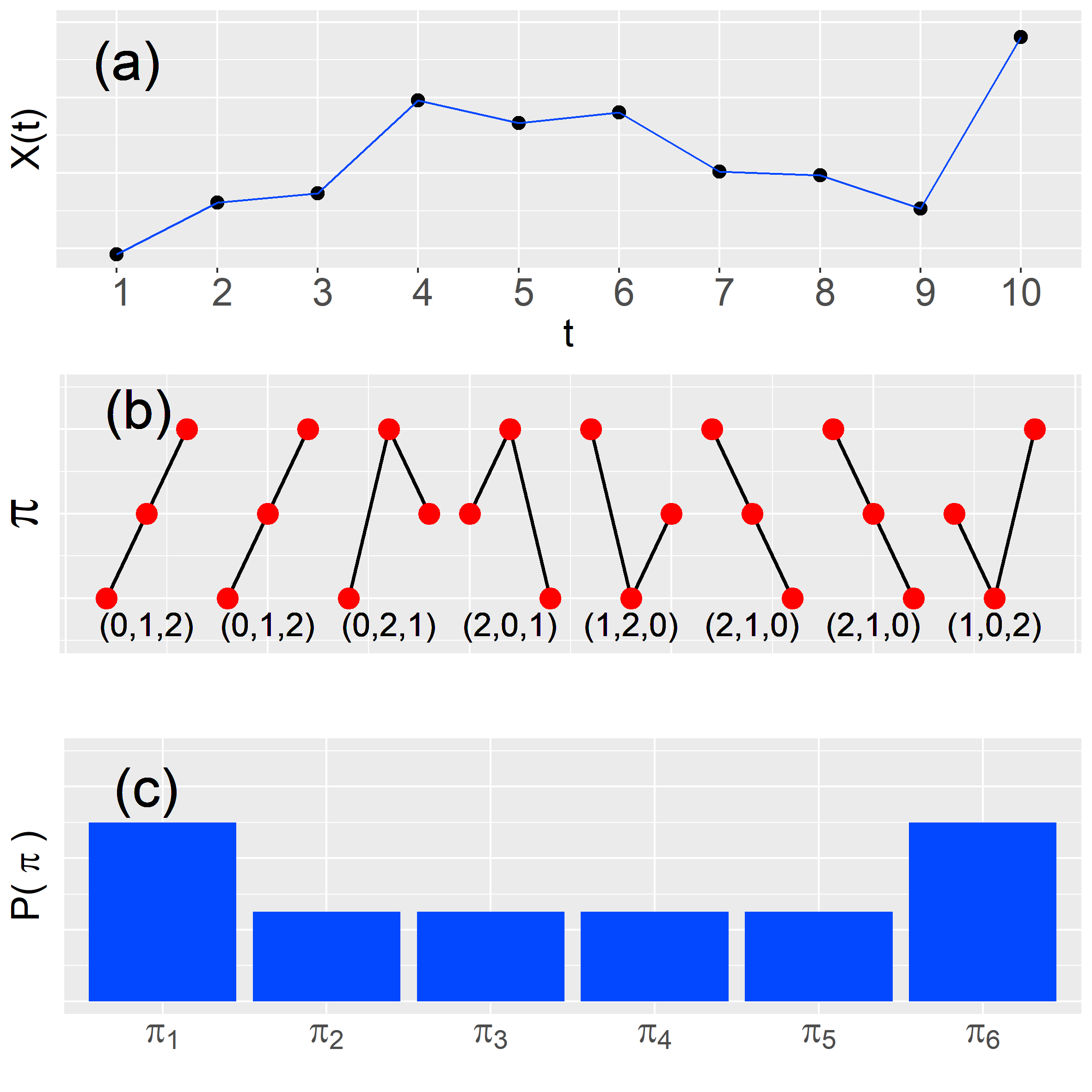}
\caption{\label{fig:patrones} 
Ordinal patterns extraction procedure for $D=3$. 
(a) Original signal $X(t)$, with length $T=9$, 
(b) Ordinal patterns extracted from $X(t)$,
(c) Ordinal patterns probability distribution (BP-PDF).}
\end{figure}

We use the Bandt-Pompe methodology ~\cite{Bandt-Pompe2002} to associate a time causality probability distribution to a time series under study. This methodology
takes into account a suitable partition based on ordinal patterns obtained by comparing neighboring series values. 
For a given time series 
${\mathcal X} \equiv \{ x_t , t = 1, \dots, T ; ~ x_t \in \mathbb{R} \} $ of length $T$, we identified the $ M = T-(D-1)$ overlapping segments
\begin{eqnarray}
X_s = ( x_{s}, x_{s+1}, x_{s+2}, \dots, x_{s+(D-1)} )  
\label{eq-vectores}
\end{eqnarray}
of length $D \in \mathbb{N}$, $D \geq 2$ (embedding dimension).
Within each segment, the component values are ordered in an increasing order to find the indices $r_0, r_1, r_2,  \dots , r_{(D-1)}$ such that
\begin{eqnarray}
x_{s+r_0} \leq x_{s+r_1} \leq x_{s+r_2} \leq \dots \leq x_{s+r_{(D-1)}} \  
\label{eq-vectores-ordenados}
\end{eqnarray}
The corresponding $D$-tuples (or words) 
$\pi = (r_0, r_1, r_2,  \dots , r_{(D-1)})$
are symbols corresponding the original segments, and can be assumed any of the $D!$ possible permutations of the set $\{0, 1, 2, \dots , (D-1) \}$.
Then the permutation entropy (Shannon entropy) of the Bandt-Pompe PDF ($\Pi^{(D)}$) is defined as
\begin{eqnarray}
S[ \Pi^{(D)} ] = - \sum_{{\{\pi}\}} p(\pi) \ln(p(\pi)) \ ,
\label{eq-entropia}
\end{eqnarray}
where ${{\{\pi}\}}$ represents the summation over all the $D!$ possible permutations of order $D$. 
$p(\pi) \geq 0$ is the relative frequency of occurrences of the permutation $\pi$ and clearly satisfied $\sum_{{\{\pi}\}} p(\pi) = 1$. 
The BP-PDF, $\Pi^{(D)}$ is invariant before monotonic transformations of the time series values.
The optimal value of the pattern length $D$ is strongly related with the phenomenology of the event under study and the availability of the data, however, as a rule of thumb, we choose the maximum $D$ value such that  $T \gg D!$ in order to obtain a good statistic. \\
Let's see an example of the procedure to obtain the BP-PDF. Figure. \ref{fig:patrones}.(a), depicts the finite collection of samples: \\
${X(t)} = \{-8.1, ~61, ~73, ~196, ~166, ~180, ~102, ~97, ~53, ~280 \}$, the time series length is $T=9$ and taken $D=3$, we can form $M=T-(D-1)=8$ segments.\\
If $D=3$, then we have 6 possible patterns, given by: 
$\pi_1 = \{ x_1 \leq x_2 \leq x_3 \}$,
$\pi_2 = \{ x_1 \leq x_3 \leq x_2 \}$,
$\pi_3 = \{ x_2 \leq x_1 \leq x_3 \}$,
$\pi_4 = \{ x_3 \leq x_1 \leq x_2 \}$,
$\pi_5 = \{ x_2 \leq x_3 \leq x_1 \}$,
$\pi_6 = \{ x_1 \leq x_2 \leq x_1 \}$.
Fig. \ref{fig:patrones}.(b), shows the eight obtained 3-tuples. \\
The pattern probability are: 
$p(\pi_1)= p(\pi_6)=2/8$ and $p(\pi_2)= p(\pi_3)=p(\pi_4)= p(\pi_5)=1/8$, which conforms the obtained BP-PDF $\Pi^{(D=3)}$ showed in Fig. \ref{fig:patrones}.(c).

\subsubsection{\label{subsec:dynamics} Dynamics}
The randomness in the system is measured trough Normalized Permutation Entropy defined as
\begin{eqnarray}
H[P] = \displaystyle{  \frac{ S[P]}{S_{max}} = \  \frac{-1}{S_{max} } {\sum_{j=1} ^{N} p_j \ln(p_j) } } \  ,
\label{eq:H-entropy}
\end{eqnarray}
where $P=\{ p_j = 1, \cdots, N \}$ denote the PDF representing the actual state of the system. $S_{max} = S[P_e] = \ln N = \ln (D!)$ is a normalization constant, obtained from the equilibrium probability distribution $P_e = \{  p_j = 1/N  ,  \forall j =  1, \cdots , N \} $. In this way, $ 0 \leq H[P] \leq 1 $, being  the extreme values for a complete ordered and uncorrelated random system. 
$H[P]$ allows comparisons between the proportion of information contained in a system with different amount of states. \\
A second measure that allows characterizing the emergence of new properties and organization in the system is the statistical complexity, such that $0 \leq C[P] \leq 1$: 
\begin{eqnarray}
C[P] = H[P] \cdot Q[P, P_e]  \ ,
\label{eq:complexity}
\end{eqnarray}
where $Q[P,P_e]$ stands for normalized Jensen-Shannon  divergence, a non-euclidean distance between observed and uniform probability distributions:
\begin{eqnarray}
Q[P,P_e] = \displaystyle{ Q_0  \left\{ S\left[\frac{(P + P_e)}{2} \right] - \frac{S[P]}{2} - \frac{S[P_e]}{2} \right\} }  \  ,
\label{eq:disequilibrium}
\end{eqnarray}
with $Q_0$ as a normalization constant ~\cite{tiana2010} 
\begin{eqnarray}
Q_0 = \displaystyle{ -2 \left\{ \left( \frac{N+1}{N} \right) \ln(N+1)  - 2 \ln(2N) + \ln(N)  \right\}^{-1} }.
\label{eq:Q0-valor}
\end{eqnarray}
\hfill\break
Note also, the global character of the both introduced quantifiers, entropy   (eq. (\ref{eq:H-entropy})) and complexity  (eq. (\ref{eq:complexity})). That is, their values do not change with different orderings of the component of the PDF. \\
The statistical complexity quantifies the existence of correlational structures, giving a measure of the complexity of a time series. 
In the case of perfect order ($H[P]=0$), or total randomness  ($H[P]=1$), of a time series coming from a dynamical system, we have $C[P]=0$, meaning that the signal possesses no structure. In-between these two extreme instances, a large range of possible stages of physical structure may be realized by a dynamical system ~\cite{Lamberti2004, LMC1995, rosso2010}. \\
So, the $C[P]$ quantifies the disorder, but also the degree of correlational structures in the time series. 
It is evident that the present statistical complexity is not a trivial function of the entropy, due to its dependence with two PDF, and it has consequences in the ranges that this information quantifier can take.
For a given entropy $H$ value, the statistical complexity $C$ runs on a precise range limited by a minimum $C_{min}$  and a maximum $C_{max}$ values. 
These extreme values depend only on the probability space dimension and, of course, on the functional form adopted for the entropy and disequilibrium ~\cite{Cotas} .\\
The temporal evolution of entropy $H[P]$ and complexity $C[P]$, can be analyzed using a two-dimensional diagram called entropy-complexity plane
$H \times C$.  
In accordance with the second law of thermodynamics, the entropy grows monotonically with time. Then, the entropy $H$ can be regarded as an arrow of time ~\cite{flechatiempo}, allowing this
form to follow the time evolution of a dynamical system or its changes of behavior with the different control parameters ~\cite{plano-parametros}. \\
The $H \times C$-plane, as a diagnostic tool, has shown to be particularly efficient at distinguishing between the deterministic chaotic and stochastic nature of time series since the quantifiers have distinctive behaviors for different types of dynamics. 
In particular, Rosso and co-workers  ~\cite{PRL-Rosso2007}  
showed that chaotic maps have intermediate entropy $H$, while complexity $C$ reaches larger values, close to those of the limit $C_{max}$. 
Moreover, similar behavior is still observed when the time series are contaminated with small or moderate amount of additive uncorrelated or correlated noise  ~\cite{HxC-ruido1,HxC-ruido2} \\
As we mention previously, ordinal patterns represent a natural alphabet from the time series and their probability distribution $\Pi^{(D)} \equiv P$ allows the computation of some measures to quantify the randomness and complexity in the related system.\\
Dynamical properties were computed for all signals. To control the experiment, each of one was contrasted with 50 surrogate versions via the Iterative Amplitude Adjusted Fourier Transform (IAAFT) ~\cite{theiler1992,dolan2001,schreiber2000}. $H\times C$-plane was generated for all dimensions $D$ to evaluate potential correlations between both parameters in all bands. Due to the intimate closeness of $\alpha$ rhythm with RS, we focused on quantifies $H$ and $C$ at cortical ROI, as well as how they are distributed at local level in the scalp for this band, in particular at an intermediate observational scale $D=5$. Remaining bands are compared at the Supplementary Material.

\subsubsection{\label{subsec: structure} Structure}
The statistical coherence $C_{ij}(f)$ quantifies the intra-frequency relation between two signals assigning $0$ ($1$) when they evolves autonomous (synchronized) at a frequency
 $f$  ~\cite{carter1987,kramer2013}.
In Eq.  (\ref{eq:coherence}), $S_{ij}(f)$ stands for the cross-spectral density between signals ${\mathcal  X}_i(t)$ and ${\mathcal  X}_j(t)$, and $S_{ii}(f)$ and $S_{jj}(f)$, for their autospectral densities, respectively.
\begin{eqnarray}
C_{ij}(f) = \displaystyle{\frac{|S_{ij}(f)|^2}{S_{ii}(f) S_{jj}(f)}} \ .
\label{eq:coherence}
\end{eqnarray}
The pondered interaction $w_{ij}$ between two signals at a specific band is taken as the average of $C_{ij}(f)$ along with its respective frequencies. The pairwise interrelation among $n$ signals unveils the connectivity structure among cortical ROIs. This connectivity is associated with the adjacency matrix $W$,
\begin{eqnarray}
W= \left\{ \begin{array}{lcc}
             w_{ij}  & i \neq j	 ,\\
             \\ 0 & {\mathrm {otherwise} } \ .
             \end{array}
   \right.
\label{eq:adjacency}
\end{eqnarray}
$W$ is a $n\times n$ positive and symmetric matrix representing the functional network of cortical ROIs, where rows and columns $i$, $j$ represent the respective nodes, and $w_{ij}$ its link weight. 
The link weight $w_{ij}$ exclusively depends on the activity correlation at certain frequencies.
The set $\{w_{ij}\}$ measures the statistical interrelations among pairwise nodes without taking into account causal relations.\\
The network science studies the way by which nodes and their links are organized. It provides a series of associated metrics allowed to quantify the network structure by taking into account their strongest links. To do that, each network is filtered by removing spurious or weakest connections. Therefore, the topological properties of the resulting graph depend on how much connectivity of the inferred network is maintained. Following an efficient cost optimization filter, we threshold the matrices by optimizing the balance between the network efficiency and its wiring cost ~\cite{fallani2017}.\\
We retrieve the minimum spanning tree of each graph by preserving its stronger links. This way, the connection density highlights the properties of the network while preserving its sparsity and maximizing the balance between the network efficiency and the link density  ~\cite{Baillet2017}. Thus we quantify network features to understand what would be the role of nodes in the process of segregation of information and to detect the ROIs of cardinal importance in the cortical wiring. Specifically, we capture the node clustering and the eigenvector centrality.\\
Clustering is defined as the capacity of a network to forms connected groups of three nodes. That is the fraction of triangles potentially connected around a node. The higher the clustering, the higher the segregation of information in the network  ~\cite{Rubinov2010}. 
The segregation is then associated with the presence of specialized processing in specific and densely regions of interconnected nodes.
Eq.  (\ref {eq:clusteringcoef}) stands for the weighted variant of the local clustering
\begin{eqnarray}
c_w(i) = \displaystyle{\ \frac{2}{k_i (k_i-1)}\sum_{j,k \in n}(w_{ij}w_{jk}w_{ki})^{1/3}  } \ ,
\label{eq:clusteringcoef}
\end{eqnarray}
that takes into account the number of links $k_i$ of a node $i$, and the weights among nodes $i, j, k$ ~\cite{onnela2005,Rubinov2010}. 
A node with high clustering is assumed to be one with a high probability of forming triads with its first neighbors, also linked among them. Otherwise, its clustering index is low. The mean clustering index represents a network that, in average, presents clustered connectivity.\\
Another role a node has is that by which it facilitates communication among specialized regions promoting a functional integration of information. Eigenvector centrality $ec(i)$ is a microscale measure that captures the node centrality by considering the global information of $W$.
While $c_w(i)$ is estimated by considering its first neighbors, $ec(i)$ made use of the importance of all nodes in the networks using spectral methods. 
$W$ is positive and real-squared, so there exist $n$ real non-negative eigenvalues. $W$ can be written as $W=UKU^T$, where $U$ is a matrix of eigenvectors and $K$ a diagonal matrix of its eigenvalues. The Perron-Frobenius theorem states that $W$ has a unique largest eigenvalue with an associated eigenvector having all positive entries. $ec$ is then the normalized version of this eigenvector, where 
$ec(i)$ is the centrality of the node $i$  ~\cite{boccaletti2006}. It implies that a node with a higher $ec(i)$ is that one that is connected to highly connected nodes, also called hubs.\\
$c_w$ and $ec$ were identified for frontal, occipital, parietal, and temporal lobes. For a better comparison, we normalized $c_w$ respect to the average of a set of 50 surrogate matrices. 
We take the mean value of clustering ${\overline {c_w}}(i)$ for all over the 40 subjects. We repeat the process by using the 50 randomized versions $c_w^*(i)$ per subject. 
The ${\overline{ c_w}}^{\, n}(i)$ is the normalized version of clustering respect to its rewired versions ${\overline{c_w}}^{\,n}(i)={\overline{ c_w}}(i) / {\overline{ c_w^*}}(i)$. 
This normalization allows us to focus on structural changes of networks, avoiding variations of the average weights of the connectivity ~\cite{Johann2018SciRep2}. 
In this vein, we keep ${\overline{c_w}}^{\,n}(i) \geq 1$, since this does not come from a random hidden structure. Normalized network features were topographically associated with each lobe, in particular for the $\alpha$ band. Finally, we performed a principal component analysis (PCA) between $\{H \wedge C\}$ and $\{c_w \wedge ec\}$ to scrutinize potential relationships between dynamics and structure for each lobe.

\section{\label{sec:} Results}
\begin{figure}
\includegraphics[width=0.5\textwidth]{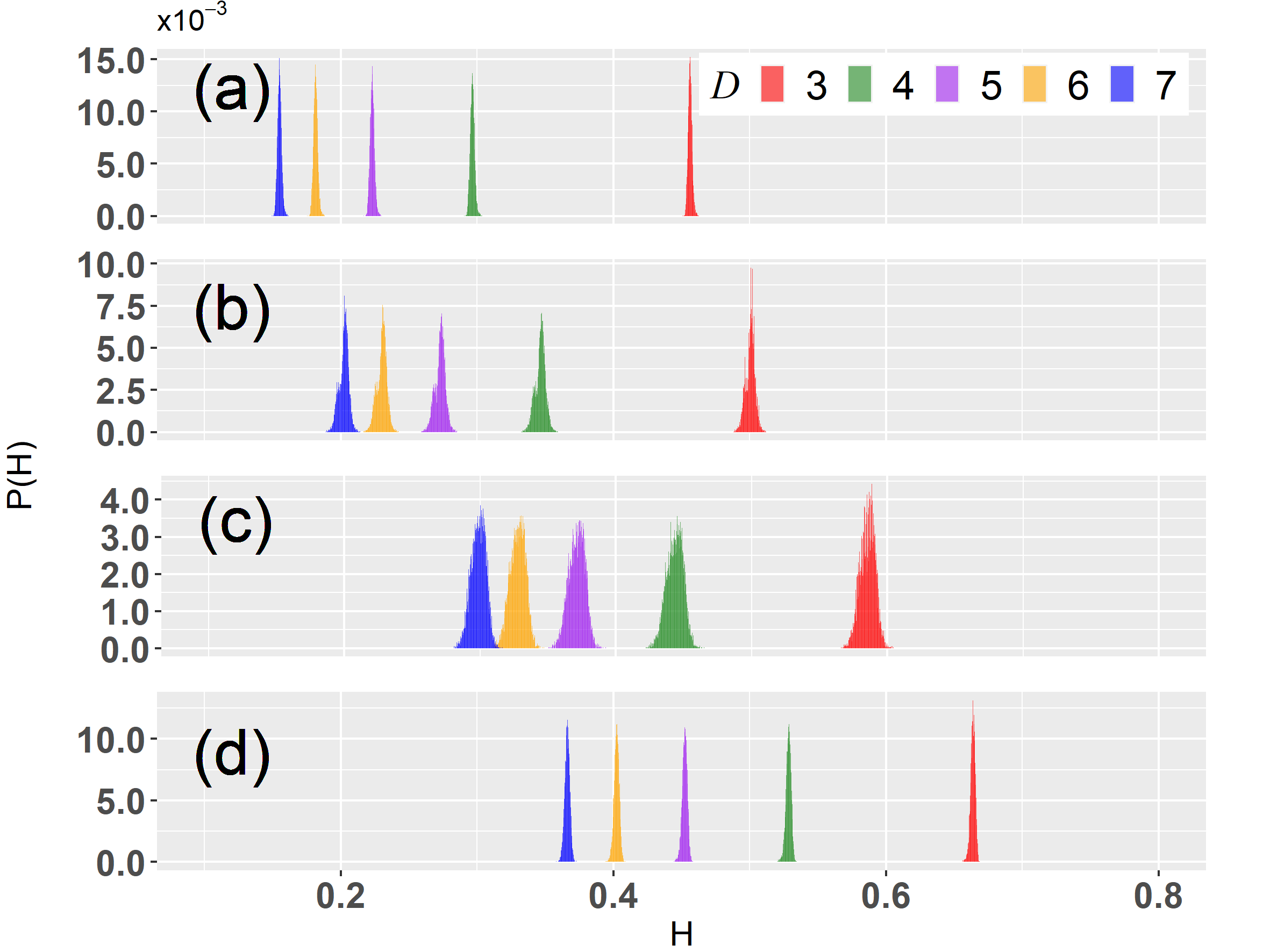}
\caption{\label{fig:entropia} Entropy probability distributions for $D=3$ (red), $D=4$ (green), $D=5$ (purple), $D=6$ (yellow), $D=7$ (blue). Each panel depicts the set of entropies per bands (a) $\theta$, (b) $\alpha$, (c) $\beta$, (d) $\gamma$. Entropy is inversely proportional to higher $D$.}
\end{figure}

We introduce results in the following way. We show the behavior of entropy $H$ and complexity $C$ for all bands at embedding dimensions $D$ ranging from 3 to 7. We make known the order relation of dynamic parameters in the $H\times C$-plane in terms of frequencies and scales. Subsequent results focused on the $\alpha$ band due to its importance for RS. We focused on the topographic distribution of $H,C$ for $D=5$. Finally, we also introduced the structural parameters $c_w$ and $ec$ for each lobe and its topographical distribution.\\
\begin{figure}
\includegraphics[width=0.505\textwidth]{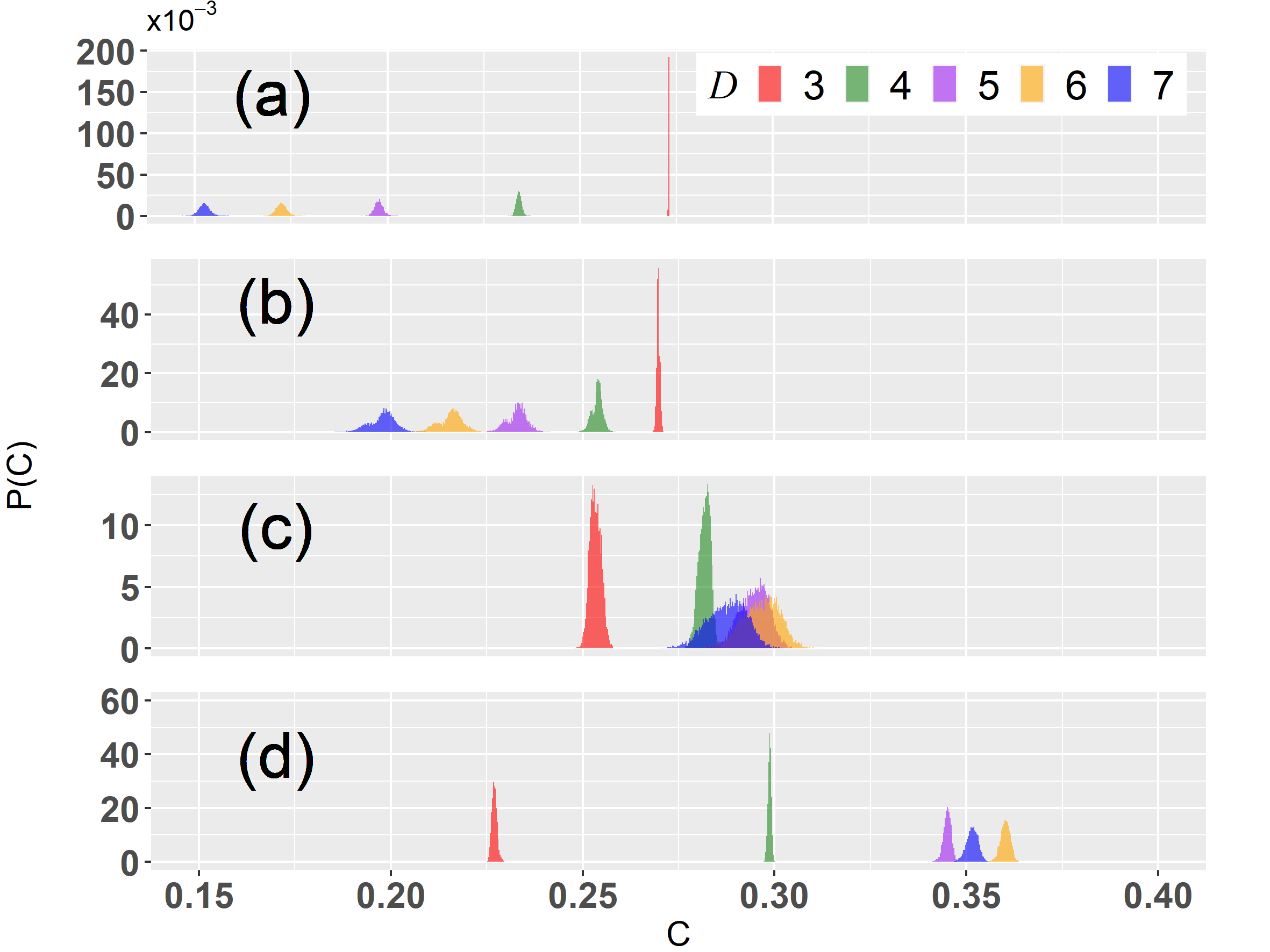}
\caption{\label{fig:complejidad} Complexity probability distributions. Similar conventions of Fig. \ref{fig:entropia}. Complexities of both (a) $\theta$ and (b) $\alpha$ decrease inversely proportional to the dimension $D$. A change of order occurs when complexities decrease for low scales $D=3,4,5$ in (c) $\beta$ and (d) $\gamma$.}
\end{figure}
Preliminary exploration of signals showed some differentiation in the autocorrelation function (ACF) in terms of frequency bands. The higher a band, the faster its ACF tends to zero. Oscillatory and sign alternation trait of ACF evidence some type of seasonal behavior. Partial autocorrelation function (PACF) takes into account inter-lag effects in contrast to ACF. The ACF of signals also evidenced a similar behavior obtained with PACF. This fact suggests that cortical dynamics might change on frequency bands for RS. 
 Although the statistical viewpoint indicates differences in bands, we are interested in the causal perspective of the signals' dynamics. In other words, we want to evidence whether such differences are captured through spatio-temporal patterns associated with each rhythm

\begin{figure*}
\includegraphics[width=1\textwidth]{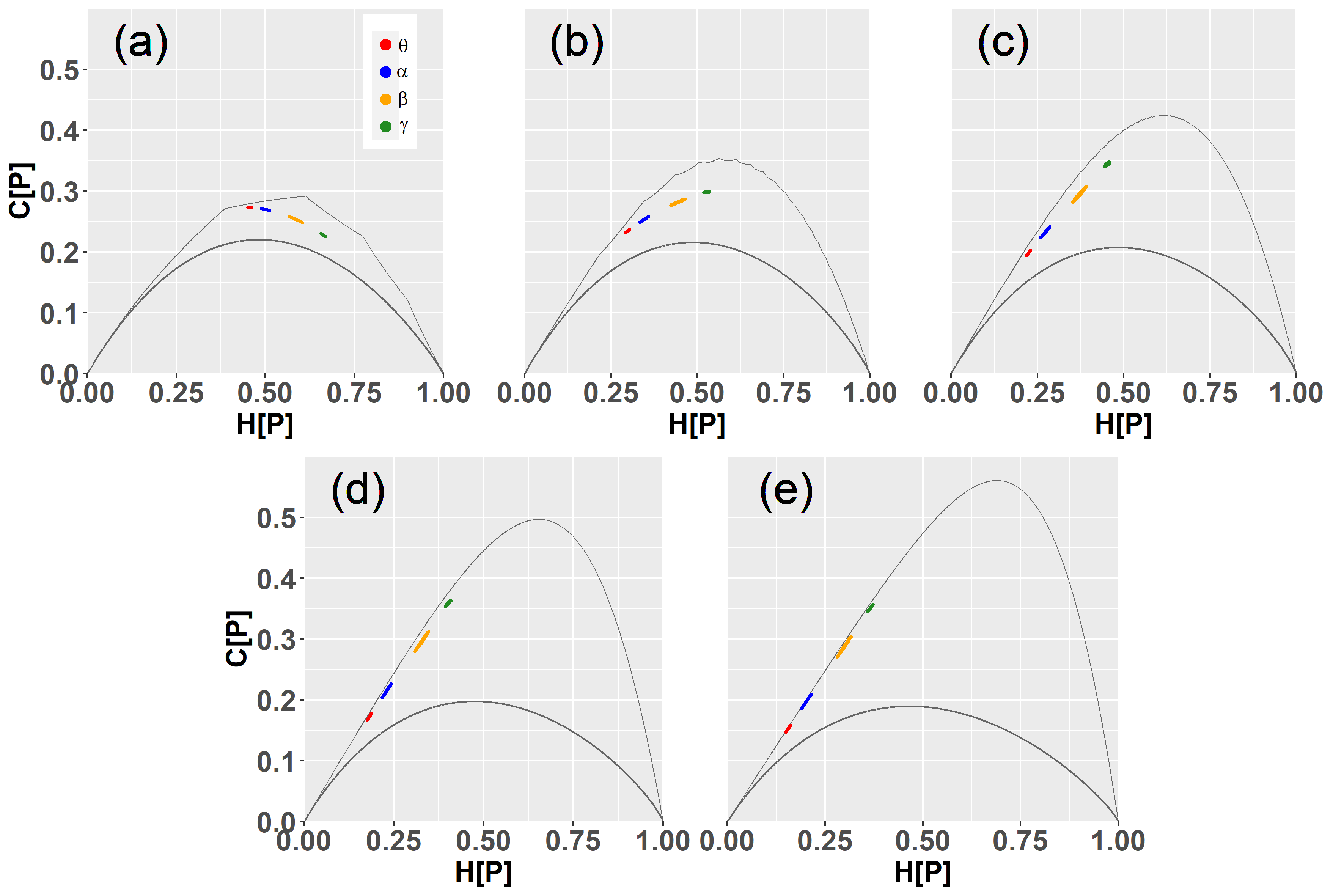}
\caption{\label{fig:hc} $H \times C$-plane for dimensions $D=3$ (a) , $D=4$ (b), $D=5$ (c), $D=6$ (d), $D=7$ (e). Each subplot accounts for bands $\theta$ (red), $\alpha$ (blue), $\beta$ (yellow), $\gamma$ (green). Gray curves represent the limits $C_{min}$ and $C_{max}$. There is an order relation among $H$ and $C$ for all frequencies.}
\end{figure*}

\subsection{\label{sec:dinamica} Dynamics}

To capture dynamical parameters, we computed the BP-PDF for filtered signals at several dimensions $D$. 
The corresponding normalized Shannon entropy $H$  and statistical complexity $C$ were evaluated. 
In Fig. \ref{fig:entropia}  and  Fig. \ref{fig:complejidad} the histogram PDF of obtained values for entropy $H$ and complexiry $C$ are shown.
For a single band, PDFs of entropy are inversely ordered respect to the dimensions $D$. The bigger the observational scale $D$, the less the entropy. 
For $\theta$ band, entropy ranges span around a order phase from ($H_{D=7}\leq0.2$) to an entropy value ($H_{D=3}\leq0.5$).
The set of entropies per dimension moves forward away from the ordered zone. In $\alpha$ band, it goes from ($H_{D=7}\approx0.2$) up to ($H_{D=3}\leq0.5$). $\beta$ band extents its entropies from ($H_{D=7}\approx0.3$) to ($H_{D=3}\approx0.58$). This pattern remains constant when exploring higher frequencies. A reduced set of symbols are allowed for $D=3$, and the richness of possible patterns have to be re-distributed in few degrees of freedom.\\
In contrast to $H$, statistical complexity $C$ behaves in a particular way. In principle, Fig. \ref{fig:complejidad}.(a) and (b), show a similar organization pattern as for $H$. The observational scale $D$ appears to be inversely ordered respect to the complexity levels. This behavior is kept in cortical activity at frequency bands $\theta$ and $\alpha$. For instance, $\alpha$ complexities spans over a range of ($C_{D=7}\approx 0.2$) up to a complexity degree ($C_{D=3}\leq0.275$).\\
The PDF for $D=3,4,5$ decrease their complexity levels for $\beta$ band, breaking it up the order previously maintained. For $\beta$, scales $D=\{3,4,5\}$, are now directly proportional to their complexity levels, for instance, $\langle C_{D=3} \rangle_{\beta}=0.253$, $\langle C_{D=4} \rangle_{\beta} = 0.281$, $\langle C_{D=5} \rangle_{\beta} = 0.294$. Cortical activity at $\gamma$ band magnifies this new ordering by expanding the distances among PDFs of $C
$ for different $D$-scales. Table \ref{mean_parameters} summarizes the mean values for $H$ and $C$ for all bands and $D$'s.\\
Previous results depict the independent organization of $H$ and $C$ based on $D$ and bands, but nothing says about a potential relationship between both parameters.
For a single value of $H$, $C$ may spans over a range limited by  $C_{min}$ and $C_{max}$. 
Then we ask for how the relationship of the entropy and complexity evolves for different bands? The entropy-complexity plane $H \times C$ allows following the time evolution of a dynamical system for different $D$ dimensions.\\
A negative relation between $H$ and $C$ occurs for $D=3$ (Fig. \ref{fig:hc}.(a)). The first bands accumulate higher levels of complexity in contrast to those of fast frequencies, which increment their randomness. While the order relation among frequencies is kept, $H$ and $C$ are now in the region of positive relationships for $D=4$ (Fig. \ref{fig:hc}.(b)). Now fast frequencies increase their complexity, while lower bands start to reach regions of deterministic behavior. This relationship increases its slope for all frequencies at $D=5$, now locate it on the deterministic region of the $H \times C$-plane (Fig. \ref{fig:hc}.(c)). This tendency is near the values of maximum complexity for $D=6$ (Fig. \ref{fig:hc}.(d)) and is very close to the limits when $D=7$. Note that although the slope of the relationship changes from negative (for $D=3$) to positive (since $D=4$) the order relation among bands is maintained for all dimensions. It may suggest the existence of a band-dynamics of its own that is well captured by the Bandt-Pompe methodology.
\begin{figure*}[htb]
\includegraphics[width=1\textwidth]{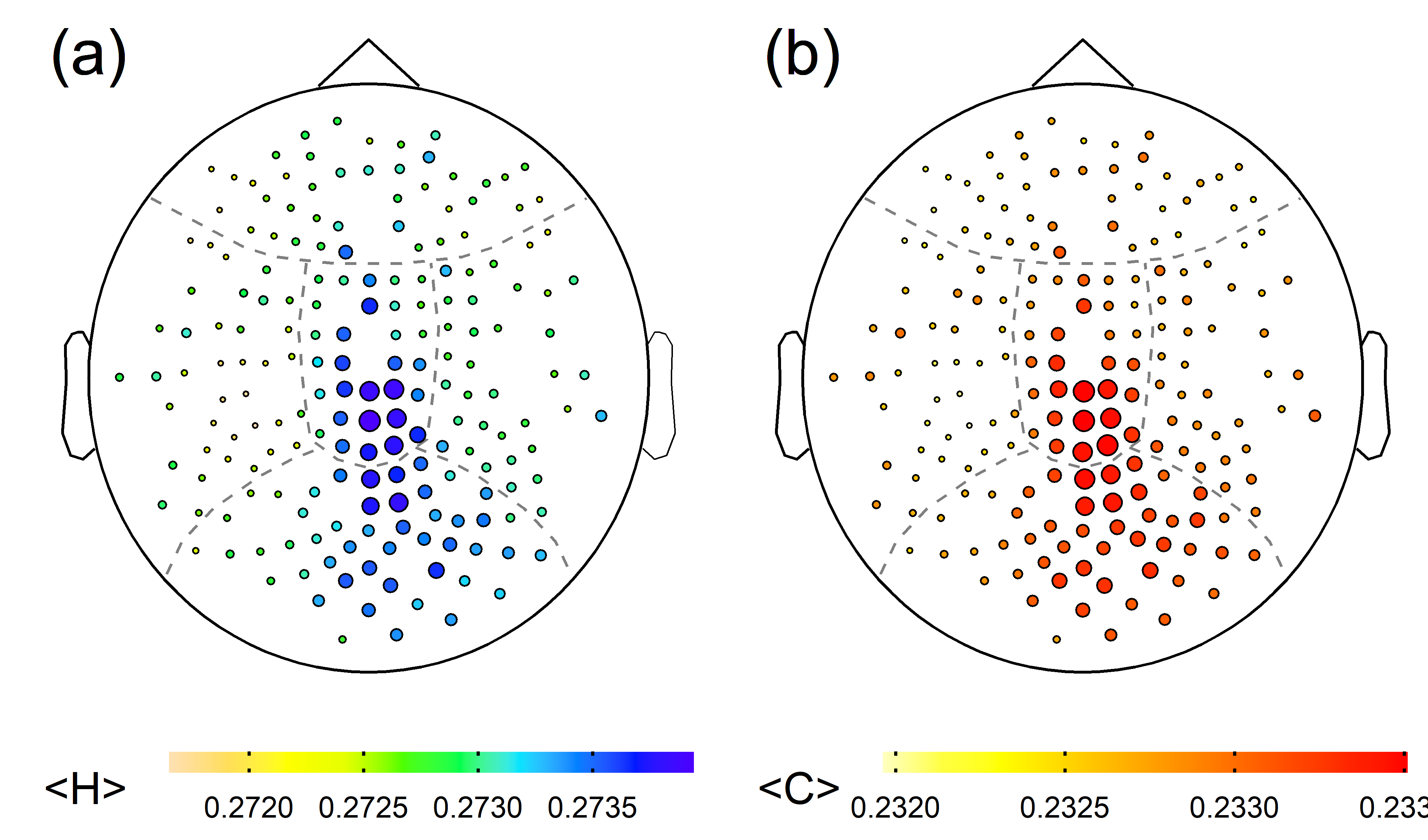}
\caption{\label{fig:hc_cabezas} Topographical distribution of $\langle H \rangle$ and $\langle C \rangle$ for $\alpha$ with $D=5$. Dashed lines divide the cortical surface into lobes. Node sizes are proportional to the average of entropy and complexity. High dynamical activity are located in posterior parietal cortex and occipital lobe. }
\end{figure*}
We chose $D=5$ to better observe its dynamics for being this the dimension in-between a short observation scale ($D=3$) and an oversampling of symbols due to $D=7$.
Under the assumption that a surrogate time series destroys all of its nonlinear dependencies, we construct a set of 50 randomized versions per signal to compute their dynamical parameters.
We compare the original $H$ (and $C$) for each ROI, respect to the empirical distribution of those captured from random versions of signals via the Kolmogorov-Smirnov test, Bonferroni corrected. 
The nonlinear character of a signal is assumed when the comparison rejects the null hypothesis that the dynamical parameters came up from time series with uncorrelated amplitudes.
Nonlinearity was detected for entropy in $\theta$, $\alpha$ and $\gamma$ oscillations with $p-value\leq0.01$. $\beta$ oscillations do not reject the null hypothesis. All bands reject the null hypothesis regarding complexity. With the previous, the information processing at the cortical level suggests to be driven by a nonlinear system presents in the dynamical properties of MEG time series.\\
Due to the relevance of $\alpha$ band for RS, we focus our attention on the topography distribution of $H$ and $C$ in the scalp so as to unveil cortical regions of high levels of dynamics. Specifically, we take the average of entropy $\langle H \rangle$ and $\langle C \rangle$ and observe their spatial distribution on scalp (Fig. \ref{fig:hc_cabezas}). Remaining bands are also showed in Fig \ref{fig:hc_all} in Supplementary Material. 
Note that $H\times C$-plane has a positive correlation for $D=5$, then nodes of higher entropy match with ROIs of higher complexity. Parieto-occipital lobe contains the majority of ROIs with high levels of $H$ and $C$. These regions exhibit an increment of cortical activity under passive attention processes and they are also associated with visual stimuli perception and spatial recognition ~\cite{huang2019}. 
These results are in accordance with RS experiments, which highlight the occipital lobe as the one with the major activity. 
In the global perspective, entropy ranges between zero and one. Its mean value for $D=5$ is rather low ($\langle H_{D=5}\rangle_{alpha}=0.272$). 
The fact that parieto-occipital lobe reaches these entropy levels may indicate that brain signals tend to have an organized distribution of its dynamics at this cortical region. 
Supplementary Material contains $\langle H\rangle$ and $\langle C\rangle$ for remaining bands in Table. \ref{mean_parameters}.

\subsection{\label{sec:estructura} Structure}
\begin{figure}
\centering
\includegraphics[width=0.48\textwidth]{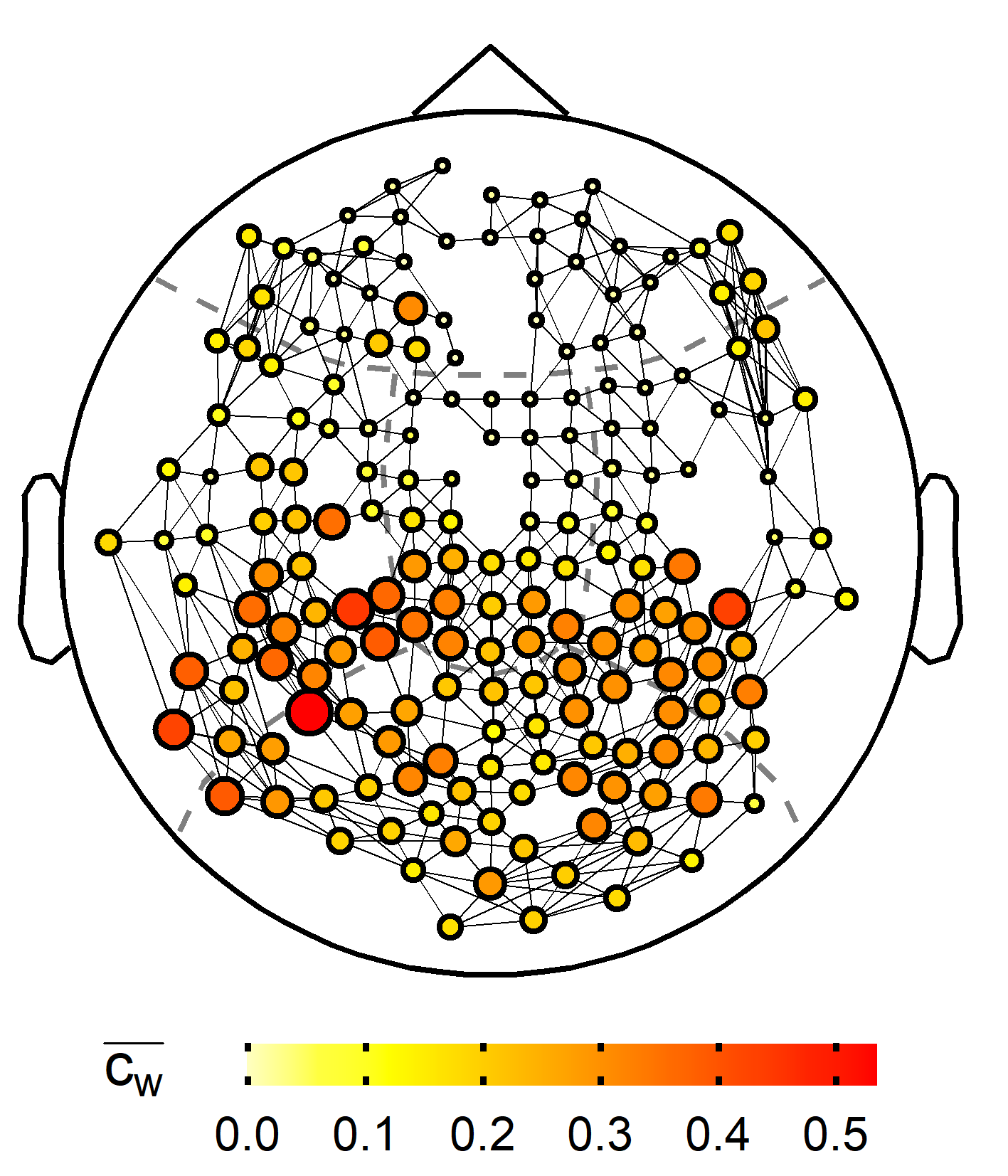}
\caption{\label{fig:red} Functional network for $\alpha$ band. Dashed lines divide cortex surface into lobes. Node sizes are proportional to clustering $\overline{c_w}$. High clustering ROIs are located mainly in temporal and occipital lobes. links are drawn with the same width for simplicity.}
\end{figure}

The dynamic analysis is performed on the activity of independent ROIs' activations. 
However, this approximation cannot take into account the statistical dependencies of ROIs' activations. 
In this sense, another spatial pattern may also arise due to these interdependencies of cortical brain signals.
Hence, we reconstructed the functional network of the $\alpha$ activity in order to capture topological features of relevance for information processing in RS.
\begin{figure}
\includegraphics[width=0.48\textwidth]{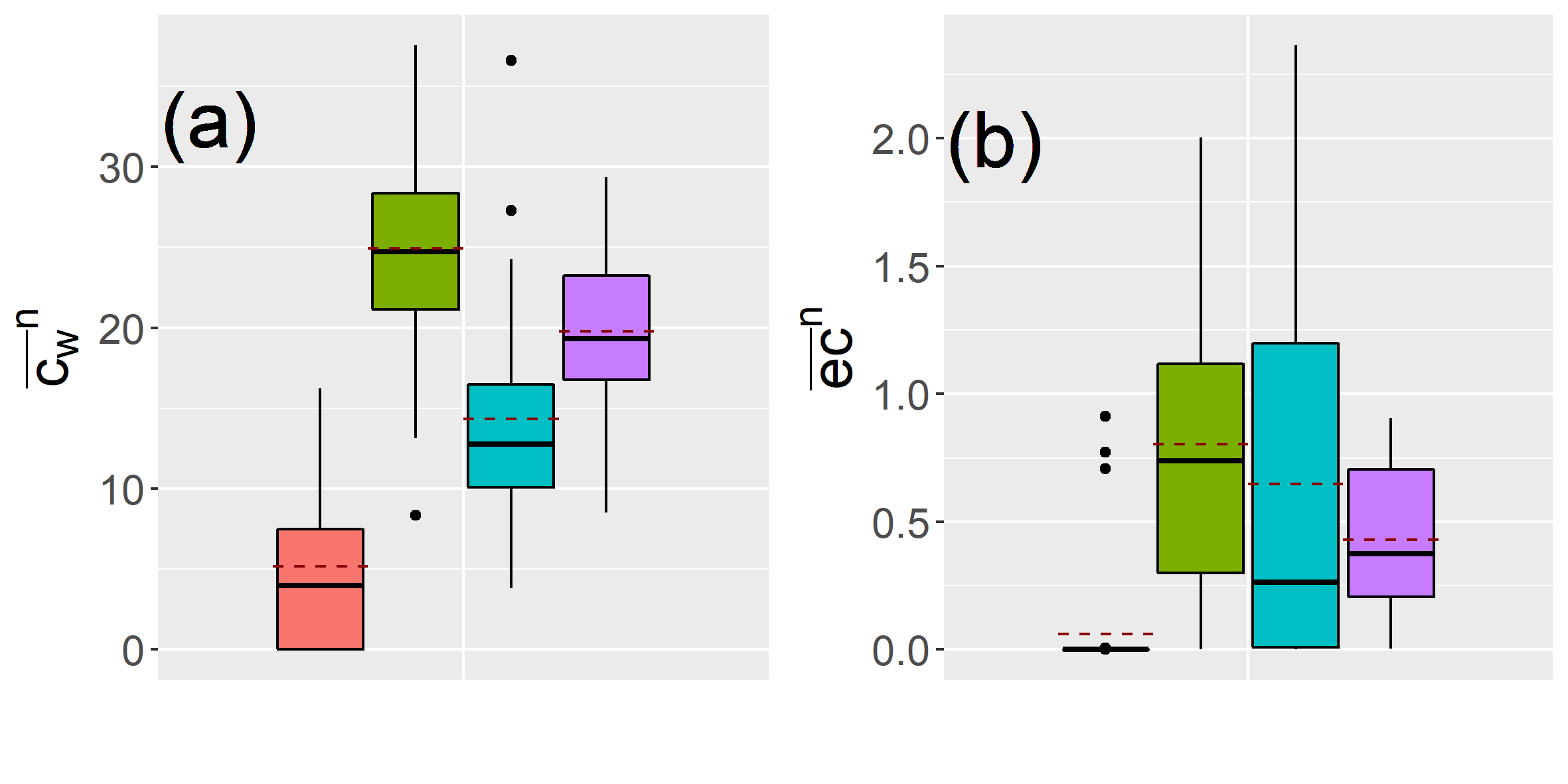}
\caption{\label{fig:parametrosnormalizados} Normalized network features. They are allocated to cortical  lobes: frontal (orange), occipital (green), parietal (cyan), temporal (purple). 
(a) Clustering ${\overline{c_w}}^{\, n}$ is higher than one for all lobes, (b) Eigenvector centrality ${\overline{ec}}^{\, n}$ is lower than one.}
\end{figure}
We computed the clustering and eigenvector centrality for all subjects and take their average value $\overline{c_w}$ and $\overline{ec}$, respectively. 
We assessing both normalized ${\overline{c_w}}^{\, n}$ and ${\overline{ec}}^{\,  n}$  by accounting for counterparts from 50 rewired networks per subject.
Network features were contrasted with those obtained from randomized matrices using a Mann-Whitnney test at the node level. Regarding the clustering index, $p-values$ were significantly less than 0.05 for occipital and posterior-parietal regions, and an important portion of the temporal lobe. There are no significant differences in eigenvector centrality. A normalized quantity higher than one implies a network parameter that is extracted from the real nature of the system. Otherwise, it would entail a network feature that is captured from a random configuration. We allocated features to each cortical lobe as it is shown in Fig. \ref{fig:parametrosnormalizados}. Boxplots of Fig. \ref{fig:parametrosnormalizados}. (a) indicate that clustering architecture arises from the structural organization of functional networks since normalized values are greater than one. Notice how the occipital lobe accumulates ROIs with the highest clustering index. Interestingly, this region also intensifies its dynamics for $\alpha$ band. On the other hand, Fig. \ref{fig:parametrosnormalizados}. (b) suggests that eigenvector are similar to those extracted from random versions of networks $\overline{ec} \leq \overline{ec^*}$.\\
Having said so, the natural step is to examine how the topographical distribution of the clustering architecture is at the node level. Fig. \ref{fig:red} depicts the functional network for $\alpha$ highlighting which nodes belong to each lobe. We highlight the nodes with a high level of clustering. They are dispersed mainly in occipital and temporal lobes, specialized regions that process information regarding attention, sensory and visual stimuli, and auditive perception ~\cite{huang2019}.
Several posterior regions of the cortex reveal ROIs with active roles for the segregation of information due to its high clustering. The fact that those regions reside in different lobes may reflect some integration type of information concerning the $\alpha$ band. The connectivity maps are shown in Fig. \ref{fig:red_all} of Supplementary Material for additional bands.

\section{\label{sec:Struct_and_dynamics} Structure and dynamics}
The occipital lobe engages the presence of high dynamics activity ($ H, C $) and a relevant role of structure ($ c_w $). For this reason, we now explore whether this lobe associates some quantifiable relationship among its dynamical parameters and network features.
A Principal Component Analysis (PCA) was performed with five variables: entropy $H$, complexity $C$, eigenvector centrality $ec$, strength $s$, and clustering $c_w$. Each variable contains 190 observations. The average along 40 subjects. We incorporate the cortical lobe as a supplementary variable related to each ROI. The factorial plane preserves the $91.6\, \%$ of the variability of the parameters. $58.4\, \%$ is in the first component and $33.2\, \%$ in the second one. See Fig. \ref{fig:biplot} at Supplementary Material. Dynamical parameters are best explained since they preserve a variability greater than $99\, \%$. See the row Total variability for $H$ (99.34), and $C$ (99.16) of Table. \ref{tabla_contribuciones}.
The network features $ec$ and $s$ are the best associated with the first component with $86.53\%$ and $66.01\%$ of the explained variability. In contrast, ($H,C$) are more related to the second component with a $59.33\, \%$ and $49.12\, \%$, respectively. 
Entropy and complexity are positively related to the second component, while network parameters are inversely related to it. This indicates linear independence between dynamics and structure.\\
Notice how Fig. \ref{fig:biplot} shows how most of the occipital and parietal nodes are located at the right-hand side of the plane, according to the results introduced above. Specifically, the occipital lobe concentrates nodes with both high dynamics and structural levels.
The fact that PCA considers a linear combination of variables and the absence of such a relationship between both classes might hint a nonlinear relationship worthwhile to explore in future works.\\
We performed a complimentary analysis. As entropy leads to complexity, and clustering accounts for significant values for all lobes, we measure the interaction of $H$ and $c_w$. We performed linear fits taken into account the node stratification for each lobe.
Linear fits are far from explaining the entropy variability throughout a linear relation with $c_w$. $R^2$ values are low. Results for linear fits are shown in Fig. \ref{fig:modelos} and 
Table \ref{models_statistics} of Supplementary Material. Nonetheless, these approximations are preliminary ones and do not imply the absence of a potential relationship. More studies must be done in order to verify any interrelation between introduced dynamical parameters and topology features.

\section{\label{sec:conclusions} Conclusions}
We have introduced the first study that characterizes the dynamics and structure of cortical fluctuations of healthy subjects in RS for different frequency bands. The cohort of this work stands out for being in an age range higher than the age of major neurodevelopmental changes, and before the onset of neurodegenerative conditions.
We use ordinal patterns as a robust methodology to extract the alphabet that constructs the information content of MEG signals. By using information theory, we capture the complexity and entropy as dynamic parameters. We found that the clustering coefficient plays an important role in the topology of the functional network created from the synchronization of the signals' powers at different frequencies. We highlight the relevance of the occipital lobe as the domain that concentrates cortical structures that play a dual role in both the dynamics and the network structure for the $\alpha$ band.\\
Ordinal patterns capture information in high-order dimensions of a nonlinear system.
The time scales at which the amount of information is captured from a dynamic system are those embedding dimensions.
In this sense, the amount of information contained in the MEG signals increases as we observe the system in low dimensionality.
The lower the dimensionality, the higher the amount of information contained in MEG signals. The larger the frequency band, the more entropy in the signals. 
However, cortical signals have entropy values below 0.6, which implies that cortical fluctuations are moderately ordered for either all observation scales and bands.
Although our study focuses on RS, it highlights that similar results were obtained at the microscale level. The work of Montani ~\cite{montani2015} developed models of spiking neurons with a diversity of different neurocomputational properties of biological neurons. Similarly, they find that the degree of order decreases as entropy increases.\\
Resting brain activity assumes an interdependence of multiple states that evolve in a non-random manner ~\cite{kenet2003,beggs2003}, which implies that there is a complex dynamic that shows the rich temporal structure of cortical activations in RS ~\cite{goldberger2002}. We then capture this complexity as another dynamic parameter of the MEG signals and associate it with their levels of randomness using the $H\times C$-plane.
Importantly, we found an order relation in the dynamics of cortical fluctuations with marked gaps between each oscillation rhythm.
The faster the oscillations, the information is processed with higher levels of complexity and entropy.\\
Echegoyen's recent work ~\cite{Echegoyen2020} discriminates MEG signals from subjects with mild cognitive impairment and Alzheimer's in different frequency bands. However, it does not show an order relation between each rhythm and the entropy ranges.
At the same time, Baravalle ~\cite{Rosso2018Chaos} demonstrated that an order relation appears for slow rhythms. However, this order relation changes for fast oscillations, specifically in EEG of subjects performing a visuomotor task.
In contrast, our results show that the $H\times C$-plane differentiates the dynamic of MEG signals with a marked order associated with each band, which implies that each rhythm has a preference to occupy specific states of the system.
Interestingly, this order relation remains independent of the observation scale, suggesting the existence of some consistency in the local dynamics associated with the RS.\\
The $\alpha$ rhythm reaches the highest power in RS and associates the posterior zone of the cortex.
In this sense, the strong dynamic of the occipital lobe is well captured by the methodology used in this work.
We evaluated the importance of cortical lobes that contain ROIs with high levels of information processing.
We find the posterior-parietal and occipital lobes as those that accumulate cortical regions where information processing associates high levels of entropy and complexity for the $\alpha$ band.
These results are consistent with those of the recent work by Quintero-Quiroz ~\cite{Masoller2018}, who analyzed the dynamics of EEG signals in RS. They found that the posterior zone of the cerebral cortex concentrates ROIs with high entropy compared to the other zones.
It is also interesting to note that this region marks the difference between the levels of entropy and complexity of MEG signals in healthy subjects with different cognitive reserves  ~\cite{Johann2018SciRep2}.\\
Overall, the methodology used discriminates spatiotemporal patterns for different frequencies.
The $\theta$ band showed a distribution in the posterior cortex and some regions of the temporal lobe that is associated with the processing of information from visual stimuli.
The $\beta$ band highlights regions of the frontal lobe, and the $\gamma$ accentuates some areas of the cortical periphery.
Interestingly, although this work focused on MEG in RS, the work of Baravalle and coworkers ~\cite{Rosso2017Entropy} highlights regions of high entropy in areas similar to those found in this work but using EEG of subjects with visuomotor tasks.
A comparison between findings allows us to find certain similarities between these two works. On the one hand, we work with MEG of RS with open eyes. On the other hand, they work on a visuomotor task of healthy subjects. Therefore, we speculate that visual activity might have a central role in the topographic distribution of ROIs with high dynamics in the different frequency bands.\\
Regarding network analysis, we find that clustering is of cardinal importance, while the role of other network parameters, such as eigenvector centrality or strength, is still unclear.
High clustering levels are due to the high density of connections between functionally related regions of the network in the $\alpha$ band.
It also is related to the phenomenon of functional segregation, in which specialized regions carry out information processing.
In this sense, we find the occipital lobe and some temporal ROIs as regions that concentrate a high level of clustering.
Interestingly, clustering has formerly been associated with the robustness of a functional network in RS.
Specifically, the clustering of functional networks is bound-up to the robustness when occurring loss of information between hemispheres in the $\alpha$ band ~\cite{Johann2018SicRep1}.
Although the work shows the importance of the occipital lobe in information processing for dynamics ($H, C$) and structure ($c_w$), no statistical relationship was found between these parameters.\\
It is well known that the MEG technique is extremely sensitive to diverse factors like skin conductance, arousal, eye movements, among others that may add noise and lead to artifacts.
For this reason, a preprocessing pipeline is continually performed for filtering and cleaning raw data.
However, some preprocessing techniques depend on both parameters that must be tuned \textit{ad hoc}, and filters that can mitigate the presence of noise but profoundly influence the results.
In this sense, although a high sampling frequency in the signals leads to a high temporal resolution, it can also induce drawbacks when capturing ordinal patterns.
An oversampling could hide the real dynamics of a signal, promoting the existence of ascending and descending patterns mostly associated with periodic time series.
The excess of these types of patterns would decrease the richness of other states that coexist in the signals, creating very ordered probability distributions that lead to low entropy.\\
Increasing the observation window increases the possible states of the system, but does not eliminate the low entropy of the distributions. As a consequence, complexity approaches the theoretical maximum levels for large embedding dimensions.
This fact may explain why the dynamical parameters in the frequency bands radically change when increasing the observation dimension.
Because the Bandt and Pompe methodology is sensitive to the observational scale, oversampling can strongly influence both the observed patterns captured and the system dynamics.
We take $D=5$ as being between a low temporal resolution observation scale and one that takes the complexity of the system to its maximum limit. This dimension has been used in previous works in neuroscience to highlight the statistical differences when comparing subjects with Alzheimer vs. healthy, or groups of different ages ~\cite{azami2015,shumbayawonda2017}.\\
Beyond being barriers, these observations can be taken as challenges useful for future studies. For example, instead of taking the amplitudes of the time series, we can consider the patterns obtained by taking the relative differences between the local maxima, when working with highly sampled signals. This procedure could avoid oversampling and capture the richness of the patterns.
On the other hand, although we did not evidence a statistical relationship between dynamics and structure, another study found a relation between entropy/complexity and clustering/strength for the cognitive reserve phenomenon.
This fact suggests the need for further studies with other perspectives to give plausible explanations about this potential relationship in RS 
(e.g., using nonlinear correlations for connectivity matrices such as information entropy or synchronization likelihood).\\
Even so, the potential of the Bandt-Pompe methodology is demonstrated independent of potential drawbacks in the signals.
The methodology has the potential to detect changes in the topographic distribution of regions with dynamic activity, as well as characterizing rhythms at which the brain operates in RS.
Although many questions remain open, to the best of our knowledge, this is one of the first studies with MEG signals that shed light on the understanding of the dynamics and structure around RS activity using a technique that involves information theory, symbolic dynamics, and network science.\\
Converging evidence suggests that Bandt-Pompe methodology is robust to noise and efficient for its ability to extract knowledge of the dynamics of a system.
In this case, we demonstrate the existence of an order relationship present in the RS dynamics and strongly marked in the frequency bands.
It should be noted that such a relationship is ubiquitous among different observational scales.
Our evidence highlights the cardinal role of the posterior region of cortex in the $\alpha$ band, particularly the occipital lobe.
This region conglomerates neural structures in charge of information processing whose play a dual role in dynamics-structure, which might be considered as a fingerprint of the RS.
\begin{acknowledgments}
J.H.M. thanks to Colciencias Call $\#811$ (Colombian Ministry of Science, Technology and Innovation) for supporting this research, and M.R. Huartos for valuable conversations. He also is grateful to J.M. Buld\'u and D. Papo for cardinal suggestions.
\end{acknowledgments}

\appendix
\section*{\label{sec:supplementary} Supplementary material}
\begin{figure*}
\includegraphics[width=1\textwidth]{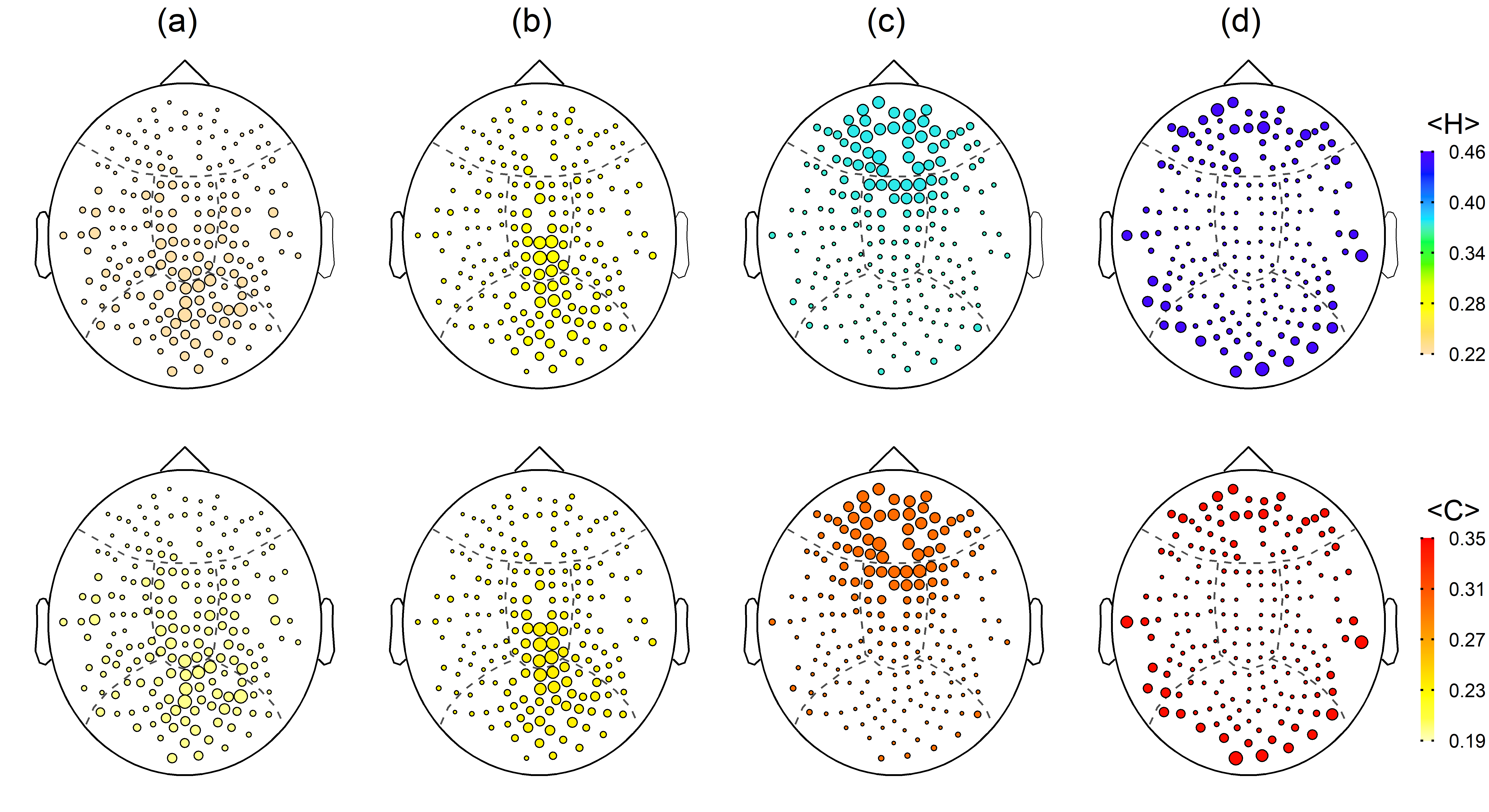}
\caption{\label{fig:hc_all} $\langle H\rangle$ and $\langle C\rangle$ for bands. Columns (a) for $\theta$, (b) for $\alpha$, (c) for $\beta$, (d) for $\gamma$. Upper panel for $\langle H\rangle$. Bottom panel for $\langle C\rangle$. Circles sizes are proportional to the dynamic parameter of each band. Color bar relates the dynamical parameters all along the frequency bands.}
\end{figure*}

\begin{figure}
\centering
\includegraphics[width=0.48\textwidth]{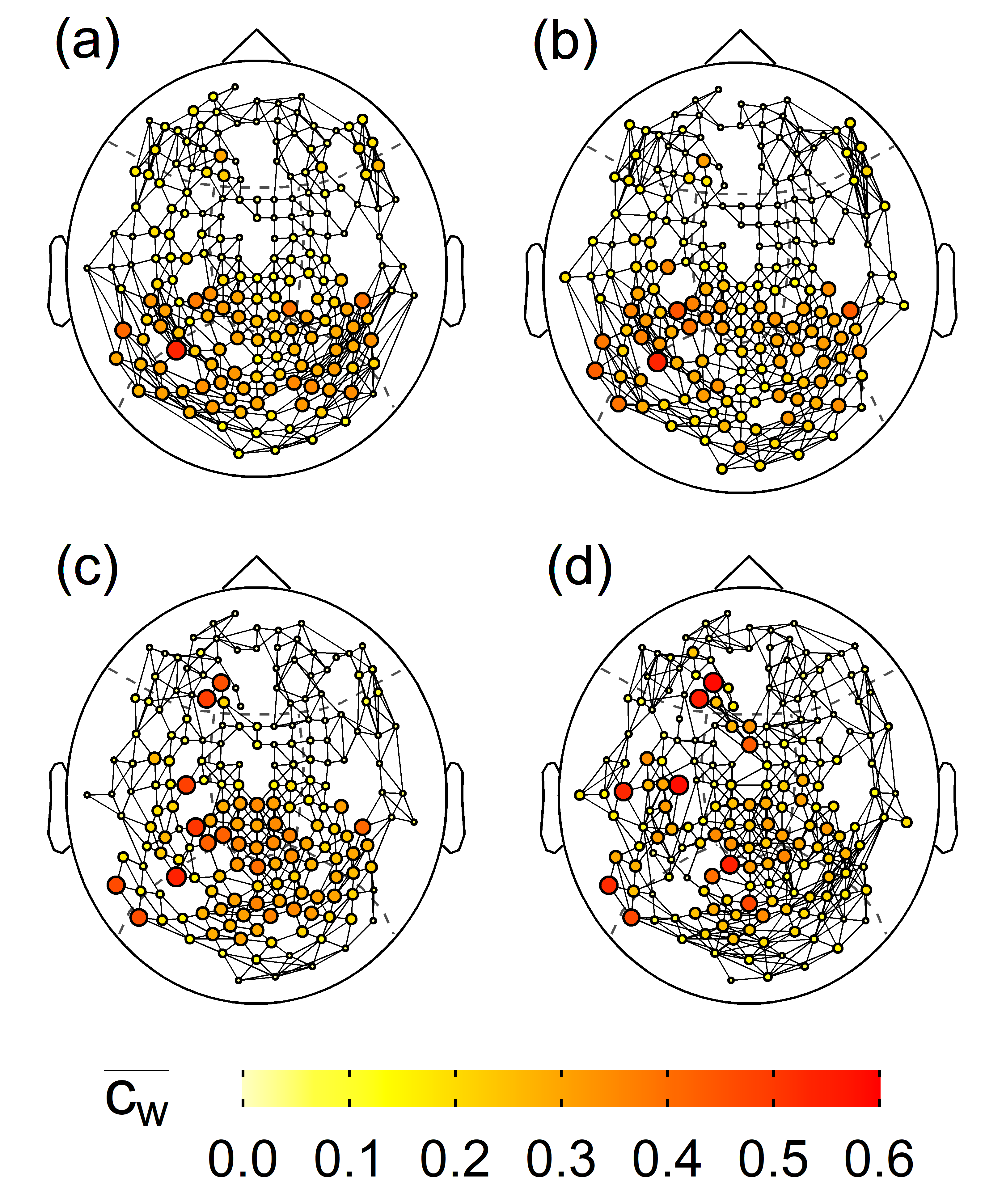}%
\caption{\label{fig:red_all} Functional networks for frequency bands. (a) $\theta$, (b) $\alpha$, (c) $\beta$, (d) $\gamma$. Node sizes are proportional to $\langle c_w\rangle$.}
\end{figure}

\begin{figure}
\includegraphics[width=0.45\textwidth]{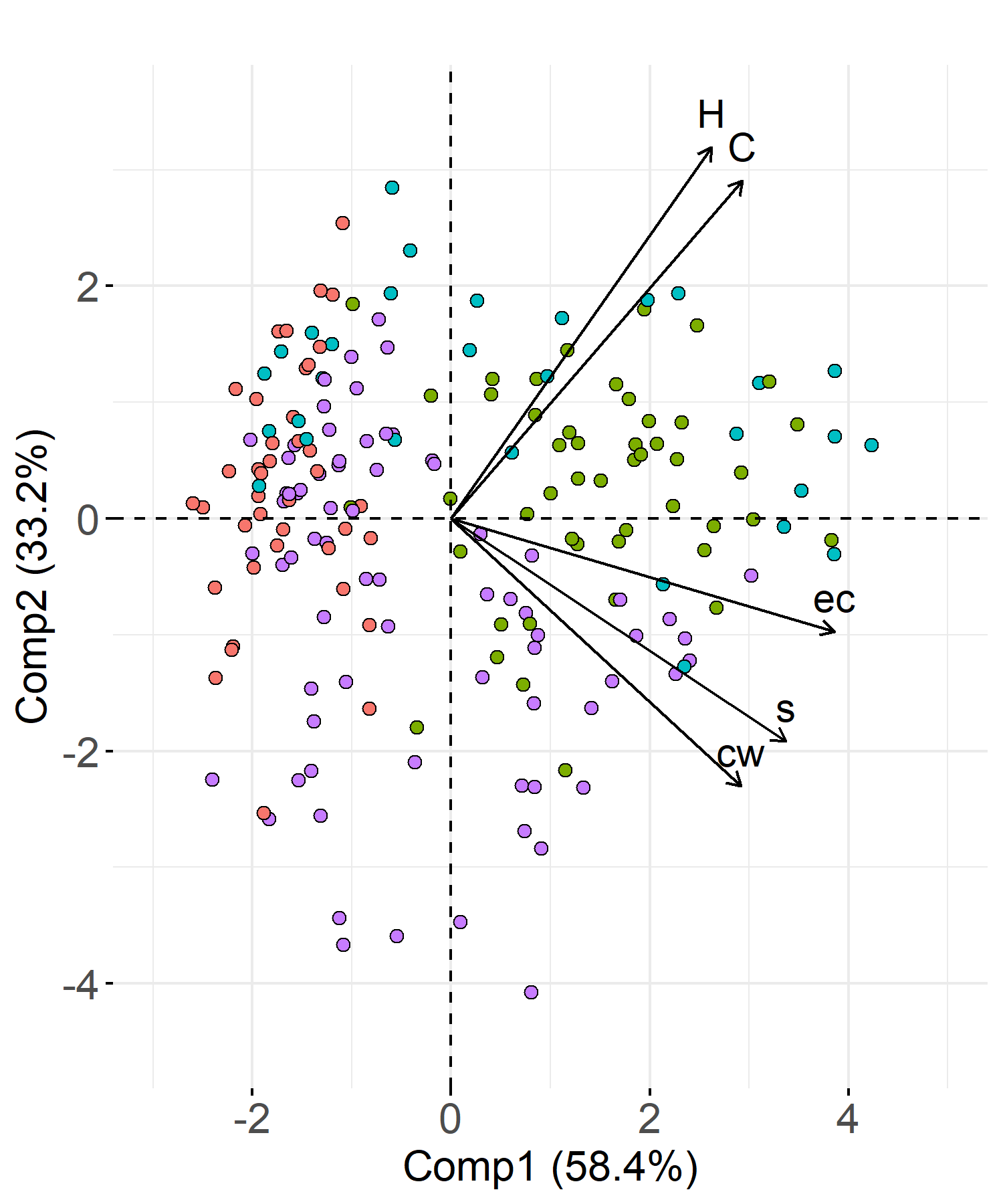}
\caption{\label{fig:biplot} Factorial plane for dynamics and structure. Frontal lobe (orange), occipital (green), parietal (cyan), temporal (purple).}
\end{figure}

\begin{figure*}
\includegraphics[width=1\textwidth]{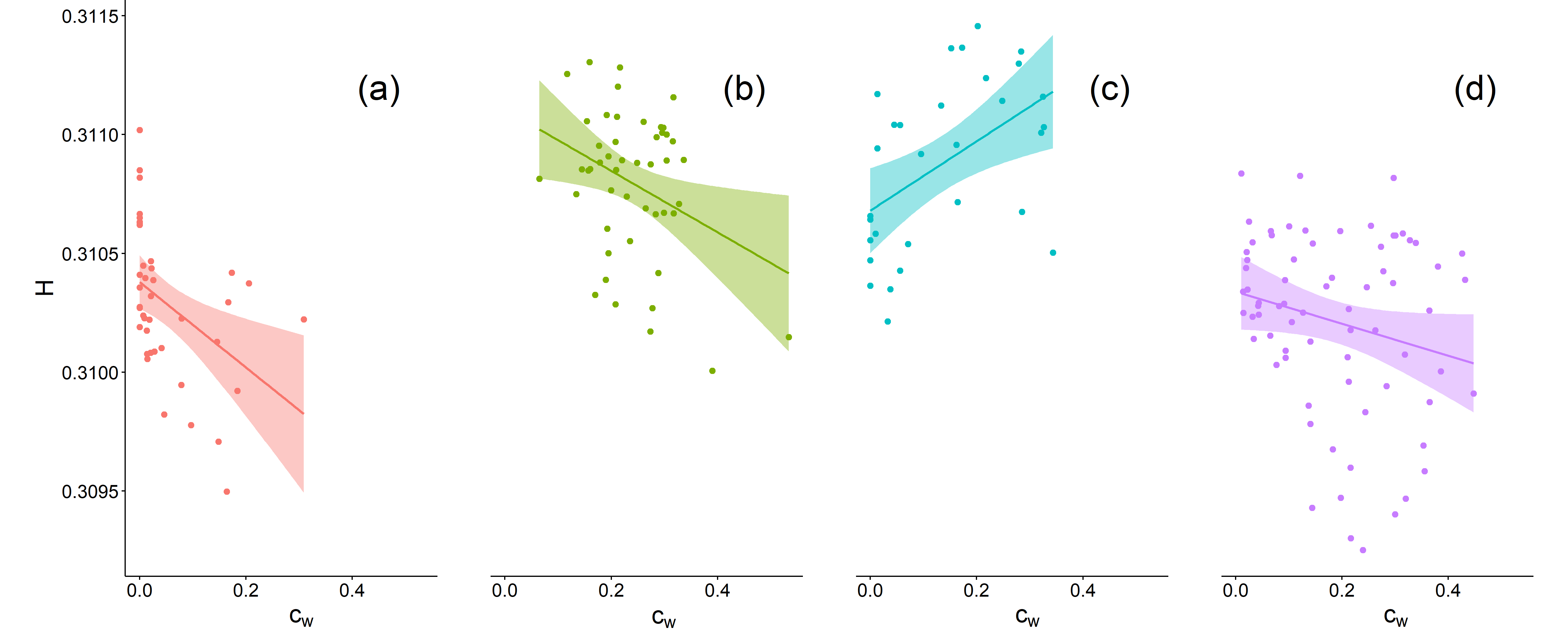}%
\caption{\label{fig:modelos} Linear fits for cortical lobes in $\alpha$. 
(a) Frontal, (b) occipital, (c) parietal, (d) temporal lobe.
 Points outside the confidence intervals show that models cannot explain the variability in entropy $H$ through with $c_w$, as explanatory variable.}
\end{figure*}

In the present work, we have focused on the dominant role of the $\alpha$ band in RS, including the adjacent frequency bands ($\theta$ and $\beta$ bands). 
Given the relevance of the $\gamma$ band (high frequencies) in resting-state for visuomotor tasks or for brain disorders, this frequency band has also been included in our analysis ~\cite{Rosso2017Entropy,huang2020,jong2018,zhou2018,andreou2014}.
This section introduces results for remaining embedding dimensions and frequency bands that were not taken into account in the main document. 
Respecting dynamics approach, the analysis was carried out considering an embedding of $D=5$ since this dimension avoids the subsampling of symbols at $D=3$ as well as the oversampling of ordinal patterns with $D=7$. In the main text, $\alpha$ band is described for being this the one that is associated with RS. Here, 
Fig. \ref{fig:hc_all} shows dynamical parameters $H$ and $C$ for remaining frequency bands. Cortical lobes are highlighted. $\theta$ band reveals higher activity the most in the occipital lobe and few sectors of the temporal lobe. These regions are associated with visual and auditory stimuli. As it was mentioned in the main text, $\alpha$ associates posterior-parietal and occipital lobes as activated regions.  
Cortical activity under $\beta$ band is located in the frontal lobe and some ROIs in the anterior parietal lobe, which is related to motor processes during RS.
$\gamma$ band focuses the activity in the outer periphery of frontal, occipital, and small regions of temporal lobes.
Table. \ref{mean_parameters} deploys mean values of $H$ and $C$ for remaining frequency bands and dimensions.
\begin{table}
\caption{Mean values $\langle H\rangle$ and $\langle C\rangle$ for bands and dimensions $D$.}
\label{mean_parameters}
\begin{ruledtabular}
\begin{tabular}{ccccccc}
Parameter & Band & \multicolumn{5}{c}{Dimension ($D$)}        
\\
\cline{1-7} 
& & $3$ & $4$ & $5$ & $6$ & $7$ \\ 
\cline{3-7}
\multirow{4}{*}{$\langle H\rangle$} & $\theta$ & 0.456 & 0.296 & 0.223 & 0.181 & 0.155 \\
& $\alpha$ & 0.500 & 0.346 & 0.272 & 0.230 & 0.202 \\
& $\beta$ & 0.585 & 0.443 & 0.371 & 0.327 & 0.298 \\
& $\gamma$ & 0.663 & 0.528 & 0.452 & 0.402 & 0.366 \\
\hline
\multirow{4}{*}{$\langle C\rangle$} & $\theta$ & 0.273 & 0.234 & 0.197 & 0.172 & 0.152 \\
& $\alpha$ & 0.269 & 0.253 & 0.232 & 0.215 & 0.197 \\
& $\beta$ & 0.253 & 0.281 & 0.294 & 0.296 & 0.287 \\
& $\gamma$ & 0.227 & 0.298 & 0.345 & 0.360 & 0.351 \\
\end{tabular}
\end{ruledtabular}
\end{table}
Regarding the architecture of functional networks, normalized features of eigenvector centrality $ec$ and weighted clustering coefficient $c_w$ were described for $\alpha$ in the main text. They appeared to be conglomerated around the occipital lobe and part of the temporal one. Here, Fig. \ref{fig:red_all} depicts the topographical distribution of clustering. 
Networks related to $\theta$ and $\alpha$ bands show a similarity in their cortical distribution of clustering. ROIs, of high functional segregation, are located at the occipital and temporal lobes. These lobes are also involved in cortical activity associated with dynamic parameters (Fig. \ref{fig:hc_all}).
Network related to $\beta$ band shows greater clustering levels in occipital lobes. Surprisingly, only the left-temporal region accumulates ROIs of high clustering. Finally, $\gamma$ band network reproduces a topographical distribution similar to $\beta$ band.
After the analysis of dynamics and structure, we focused our attention on inspecting a potential relationship between them. Firstly, we bet for a PCA for dynamic parameters $\{H,C\}$, and structural features $ec, s, c_w$.
Table. \ref{tabla_contribuciones} describes the first and second components of PCA. They explain less than $100\%$ of the variability related to dynamical parameters, and more than $80\%$ from topological features. The first of two components explain the $91.6\%$ of the variability, see the axis of Fig. \ref{fig:biplot} that shows the factorial plane of PCA.
\begin{table}
\caption{Variability $\%$ explained by PCA.}
\label{tabla_contribuciones}
\begin{ruledtabular}
\begin{tabular}{cccccc}
\multirow{2}{*}{Component} & \multicolumn{5}{c}{Parameter}        \\ 
\cline{2-6} 
			        & $H$     & $C$     & $ec$  & $s$   & $c_w$    \\ \hline
$1^{st}$			& 40.01   & 50.04   & 86.53 & 66.01 & 49.54  \\ 
$2^{nd}$			& 59.33   & 49.12   & 5.48  & 21.30 & 30.76  \\ 
Total				& 99.34   & 99.16   & 92.01 & 87.31 & 80.30  \\ 
\end{tabular}
\end{ruledtabular}
\end{table}
In Fig. \ref{fig:biplot}, sensors located in the occipital lobe are colored by green. They are located on the right-hand side of the first component. Most of the parietal ROIs (cyan) are also located in the same sector of the factorial plane. Both lobes concentrate cortical regions of high dynamics and clustering in accordance with the direction of vectors, which are disposed of respect to the positive part of the first component. Sensors at the frontal lobe do not exhibit any relationship with dynamics or structural parameters.
In a second approximation, we performed linear fits between the entropy and clustering index. Fig. \ref{fig:modelos} shows linear fits for all lobes in the $\alpha$ band. Each lobe accounts for a canonical $y=b+m \, x$ model of its own. For instance, the occipital lobe accounts for: $H=0.31-0.0013  \, c_w$.
Table. \ref{models_statistics}, introduces the statistics of linear models performed on each lobe.
Models do not explain enough entropy variability ($R^2$).
\begin{table}
\caption{Linear model statistics between $H$ and $c_w$ for each lobe.}
\label{models_statistics}
\begin{ruledtabular}
\begin{tabular}{ccc}
\multirow{2}{*}{Region} & \multicolumn{2}{c}{Statistic}        \\ 
\cline{2-3} 
                            & $R^2$ & \textit{p-value} \\ \hline
Frontal			& 0.19 & 0.005 \\                            
Occipital                          & 0.11 & 0.002\\
Parietal                          & 0.24 & 0.005\\
Temporal                          & 0.04 & 0.063\\
\hline
\end{tabular}
\end{ruledtabular}
\end{table}

\section*{Data availability}
The experimental MEG data set correspond to ``Human Connectome Project”.
See detailed explanations in Van Essen's,  and Larson's work ~\cite{van2012,larson2013}. 
The data that support the findings of this study are available within the article.


\bibliography{aipsamp2}

\end{document}